\documentclass[prb,twocolumn,showpacs,floatfix]{revtex4-1}
\usepackage{graphics,epsfig,amsfonts,amssymb,amsmath,soul}
\begin{document}
\title{Magnetic domain wall dynamics in wide permalloy strips}
\author{Virginia Est\'evez}
\email{virginia.esteveznuno@aalto.fi}
\author{Lasse Laurson}
\affiliation{COMP Centre of Excellence and Helsinki Institute of Physics,
Department of Applied Physics, Aalto University, P.O.Box 11100, 
FI-00076 Aalto, Espoo, Finland.}
\begin{abstract}
Domain walls in soft permalloy strips may exhibit various equilibrium 
micromagnetic structures depending on the width and thickness of the strip,
ranging from the well-known transverse and vortex walls in narrow and thin
strips to double and triple vortex walls recently reported in wider
strips [V. Est\'evez and L. Laurson, Phys. Rev. B {\bf 91}, 054407 (2015)].
Here we analyze the field driven dynamics of such domain walls
in permalloy strips of widths from 240 nm up to 6 $\mu$m,
using the known equilibrium domain wall structures as initial configurations.
Our micromagnetic simulations show that the domain wall dynamics in  
wide strips is very complex, and depends strongly on the geometry of the 
system, as well as on the magnitude of the driving field. We discuss in 
detail the rich variety of the dynamical behaviors found, including dynamic 
transitions between different domain wall structures, periodic dynamics of 
a vortex core close to the strip edge, transitions towards simpler domain wall structures of the multi-vortex domain walls controlled by vortex polarity, and the fact that for some combinations 
of the strip geometry and the driving field the system cannot support a 
compact domain wall.  
\end{abstract}
\maketitle

\section{Introduction}
 
The study of static and dynamic properties of the magnetic domain walls 
(DWs) in ferromagnetic nanostructures has recently attracted a lot of 
attention not only due to the related fundamental physics aspects, but 
also because of the emergence of technological applications based on DWs 
and their dynamics, such as memory\cite{parkinracetrack,barnes2006} and 
logic devices\cite{cowburn2000,allwood2002,allwood2005}. Consequently, DW 
dynamics in nanostrips and wires has been extensively studied, using both 
applied magnetic fields\cite{ono99,atkinson2003,beach2005,nakatani2005,
reviewthiaville,hayashi2007,yang2008,moriya2010, ZIN-11,hayashi2008} and 
spin-polarized electric currents\cite{spcklaui2003,thiaville2004,klaui2005,
vernier2004,thiaville2005,yamaguchi2004} as the driving force. 

One of the typical materials used for the study of DWs and their dynamics 
in strip geometry is permalloy, a soft magnetic material with a negligible 
magnetocrystalline anisotropy\cite{buschowbook,coeybook}. Such systems 
exhibit in-plane domains along the long axis of the strip, induced by 
shape anisotropy. Previous works have shown that the structures of the 
DWs separating such domains may be relatively complex, and depend strongly 
on the sample geometry\cite{nakatani2005,reviewthiaville,McMichael97,klaui2004,
nosotrosequilibrium}. For nanostrip geometries, with the strip width and thickness
of the order of 100 nm and 10 nm or less, respectively, the known equilibrium 
DW structures are the transverse DW (TW) and the asymmetric transverse DW 
(ATW)\cite{klaui2004,atwbackes2007,JAM-14,nozaki99,prbklaui2003,
annularklaui2006,vortexklaui2006}. For somewhat wider and/or thicker 
strips, the vortex DW (VW) is the stable structure\cite{klaui2004,nozaki99,
prbklaui2003,annularklaui2006,vortexklaui2006,hempe2007}. A related,
three-dimensional vortex-like flux-closing DW structure has been reported 
in permalloy strips of thickness exceeding 60 nm.\cite{nguyen2015}
Recently we have shown that in the case of even wider permalloy strips than
those with VW as the equilibrium DW structure, two additional equilibrium 
DW structures appear, namely the double vortex (DVW) and triple vortex walls 
(TVW)\cite{nosotrosequilibrium}.      

Studies of field and current driven dynamics of DWs in permalloy strips 
have focused mostly on narrow strips with either TW or VW as the equilibrium 
DW structure\cite{nakatani2005,reviewthiaville,beach2005,hayashi2007,yang2008,
moriya2010,lee2007,weerts2008,kunz2006,tretiakov2008, clarke2008,martinez2009}. 
In both cases, the DW dynamics exhibits a Walker breakdown\cite{walker74}, 
an instability occurring when the DW internal degrees of freedom are 
excited by a strong enough external driving force. This force can be a magnetic field $B>B_\text{W}$ or a spin-polarized electric current 
$J>J_\text{W}$, with $B_\text{W}$ and $J_\text{W}$ the Walker field and
current density, respectively\cite{beach2005,hayashi2007,glathe2008,
hayashi2008}. For small driving forces, the DW velocity increases
up to $B_\text{W}$ or $J_\text{W}$ (steady, or viscous regime). At the
onset of the Walker breakdown, the DW velocity decreases abruptly as a 
consequence of periodic transitions between different DW structures.
In narrow and thin strips with TW as the equilibrium structure, repeated 
transitions between TWs of different polarities (or signs of the TW 
internal magnetization) take place via nucleation and propagation of an 
antivortex across the strip width; we will refer to the transient state
as the antivortex wall, or AVW\cite{reviewthiaville}. For wider strips 
with a VW equilibrium DW, periodic transitions between VW and TW
structures are typically observed\cite{reviewthiaville,lee2007}.
However, so far little is known about the DW dynamics in even 
wider strips, especially ones with DVW or TVW as the equilibrium DW 
structure.  

\begin{figure}[t!]
\leavevmode
\includegraphics[trim=1cm 0cm 1cm 2cm,clip,width=0.9\columnwidth]
{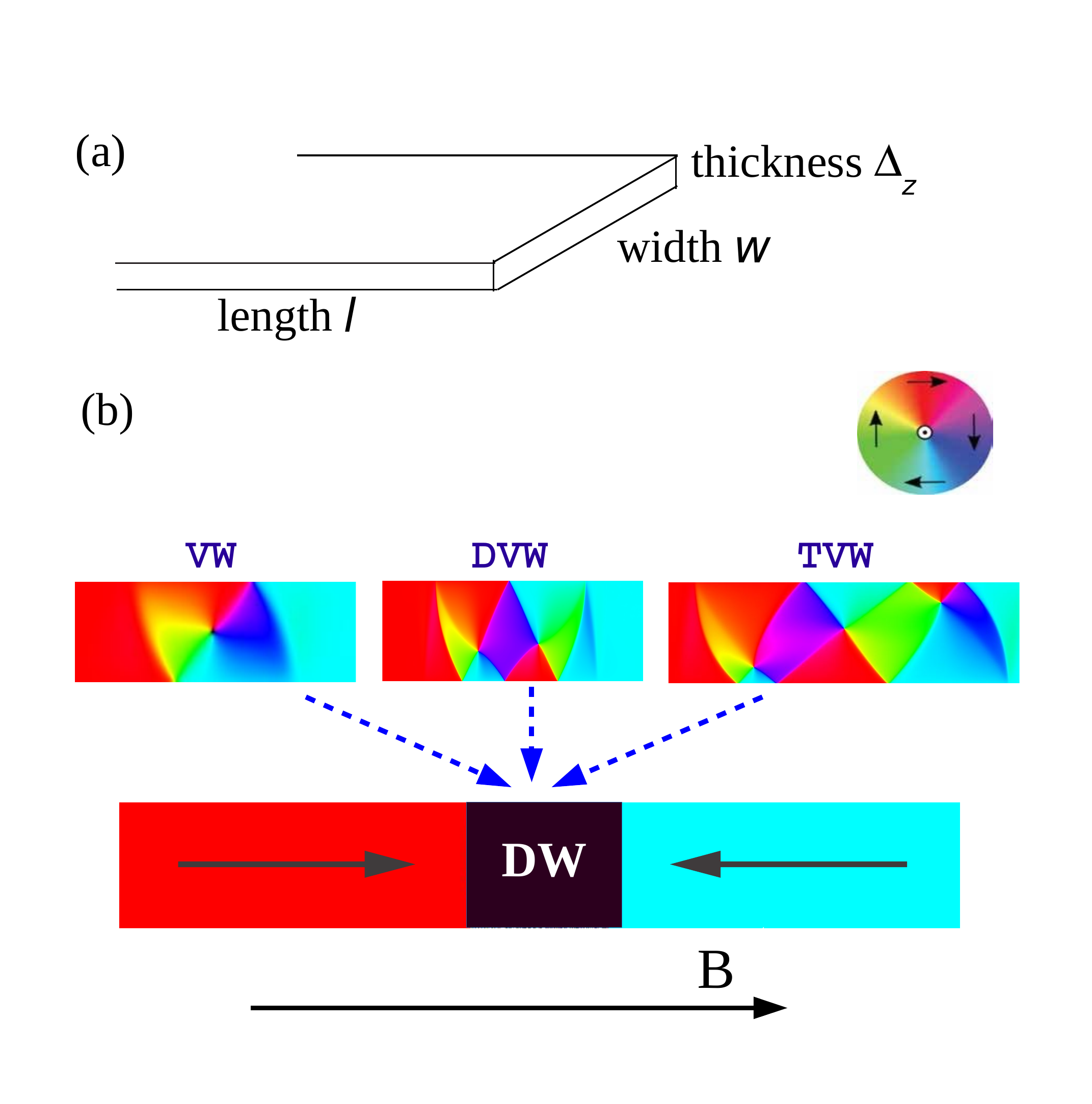}
\caption{(color online) (a) Geometry of the permalloy strip. 
(b) A top view of the magnetization in the initial state. Magnetization points 
along the long axis of the strip within the two domains (as indicated by the 
arrows) forming a head-to-head configuration. The domain wall separating these
domains has a geometry-depedendent equilibrium structure. Here, we consider
systems where the latter is either a vortex wall (VW), double vortex wall (DVW), 
or a triple vortex wall (TVW). To study the DW dynamics, an external magnetic 
field $B_\text{ext}$ is applied along the long axis of the strip.}
\label{fig:fig1_dynamics}
\end{figure}

In this paper we analyze numerically field driven DW dynamics in permalloy
strips considering a wide range of strip widths $w$ from 240 nm 
up to 6 $\mu$m. We focus on relatively thin strips with the thickness $\Delta_z$ 
in the range of 5 to 25 nm. Depending on $w$ and $\Delta_z$, the equilibrium 
DW structure in such strips can be either VW, DVW or TVW\cite{nosotrosequilibrium}
(see Fig. \ref{fig:fig1_dynamics} for examples of these structures). 
Since the space of parameters spanned by $w$, $\Delta_z$ and $B_\text{ext}$ is 
quite large, we focus on specific example cases, covering each of the three 
above-mentioned equilibrium DW structures, and illustrating 
the complex nature of the DW dynamics in wide strips, exhibiting also a strong 
dependence on both the geometry of the system, and on the magnitude of the driving 
field. Already for rather confined geometries with VW as the equilibrium DW
structure, several possible dynamical behaviors are encountered, in particular 
for large enough $B_\text{ext}$ (i.e. above the small field steady/viscous regime). 
In addition to various periodic transitions occurring between different DW structures, 
we also report on localized oscillatory motion of the vortex core close to the
strip edge. This novel behavior is characteristic of wide enough strips, giving rise 
to an extended plateau in the velocity vs field curve. Another signature
of DW dynamics in wide strips is that for some combinations of $w$, 
$\Delta_z$ and $B_\text{ext}$, DWs are unstable: the DW's  width 
grows without limit as different components of the DW assume a different
propagation velocity\cite{ZIN-11}. DVWs are found to be structurally stable
against small applied fields depending on the polarity of the vortices. In the stable case (parallel polarity), the dynamics exhibits a unique small field 
steady/viscous regime, but larger fields tend to induce a transition
into a simpler VW structure, or lead to complex, turbulent-like dynamics 
with repeated transitions between numerous different DW structures. For antiparallel polarity, the DVW transforms in a VW even for very small fields. For larger fields the dynamics observed is the same independently of the polarity. The same is 
true also for TVWs, whose stability at low fields depends on the polarities of the three vortices. In general the TVWs tend to transform into simpler DW structures
during their field-driven dynamics. The paper is organized as follows: The
following section (Section \ref{micromagn}) describes the details of our
micromagnetic simulations, while in the three sections after that (Sections \ref{vortex},
\ref{doublevortex} and \ref{triplevortex}), DW dynamics starting 
from VW, DVW and TVW equilibrium initial configurations, respectively,
are described and discussed. Section \ref{summary} finishes the paper with
a summary and conclusions.

\section{Micromagnetic Simulations}
\label{micromagn}

The system under study consists of a permalloy strip of width $w$ and 
thickness $\Delta_z$, satisfying $\Delta_z \ll w$, see Fig.~\ref{fig:fig1_dynamics} (a). 
To mimic an infinitely long strip, the magnetic charges at the two ends of the strip
are compensated. In all the cases considered the actual simulated length $l$, 
i.e. the length of the computional window which is centered at and moving with the DW, satisfies $l \geq 4w$. The initial state 
is an in-plane head-to-head domain structure, with the equilibrium DW 
structure\cite{nosotrosequilibrium} corresponding to the $w$ and $\Delta_z$ of the 
strip in the middle of the sample, see Fig.~\ref{fig:fig1_dynamics} (b). 
To analyze the DW dynamics, a constant external magnetic field $B_\text{ext}$ is 
applied along the long axis of the strip. All the results presented in this work 
have been obtained using the typical material parameters of permalloy, i.e. 
saturation magnetization $M_\text{s}=860 \times 10^3$ A/m, exchange constant 
$A_\text{ex} = 13 \times 10^{-12}$ J/m, and the Gilbert damping constant $\alpha=0.01$. 
We focus on the ideal case of perfect strips without any kind of disorder or 
impurities, and consider a temperature $T$ equal to zero.

The micromagnetic simulations have been performed using the GPU-based micromagnetic 
code MuMax3 \cite{mumax3,mumax2011,mumax2014}. To calculate the magnetization dynamics 
of the system, the Landau-Lifshitz-Gilbert equation \cite{gilbert2004,brownmicromagnetism},
\begin{equation}
\partial {\bf m}/\partial t =
\gamma {\bf H_{eff}} \times {\bf m} + \alpha {\bf m} \times
\partial {\bf m}/\partial t,
\end{equation}
is solved numerically. Here, ${\bf m}$ is the magnetization, $\gamma$ the 
gyromagnetic ratio, and ${\bf H}_\text{eff}$ the effective
field, with contributions due to exchange, Zeeman, and demagnetizing
energies. The size of the discretization cell used depends on the system 
size, ranging from 3 to 5 nm, but is always bounded by the exchange length, 
$\Lambda = (2A/\mu_0 M_s^2)^{1/2}\approx 5$ nm, in the in-plane directions, and 
equals $\Delta_z$ in the out-of-plane direction. The DW velocity is calculated  once the DW structure is in the steady state; thus, the total simulation time considered is at least 300 ns.

\begin{figure}[t!]
\leavevmode
\includegraphics[clip,width=0.5\textwidth]{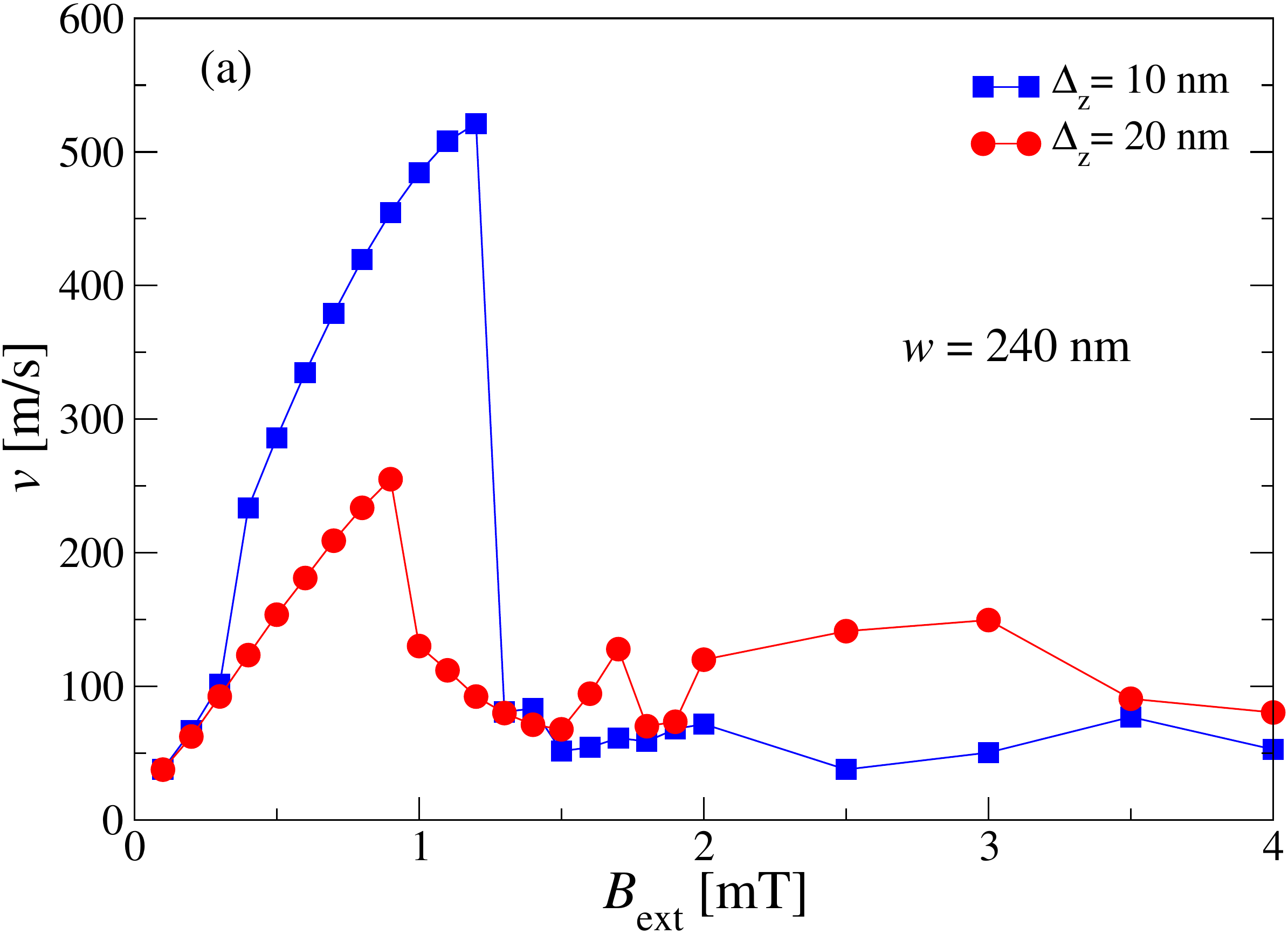}
\includegraphics[clip,width=0.5\textwidth]{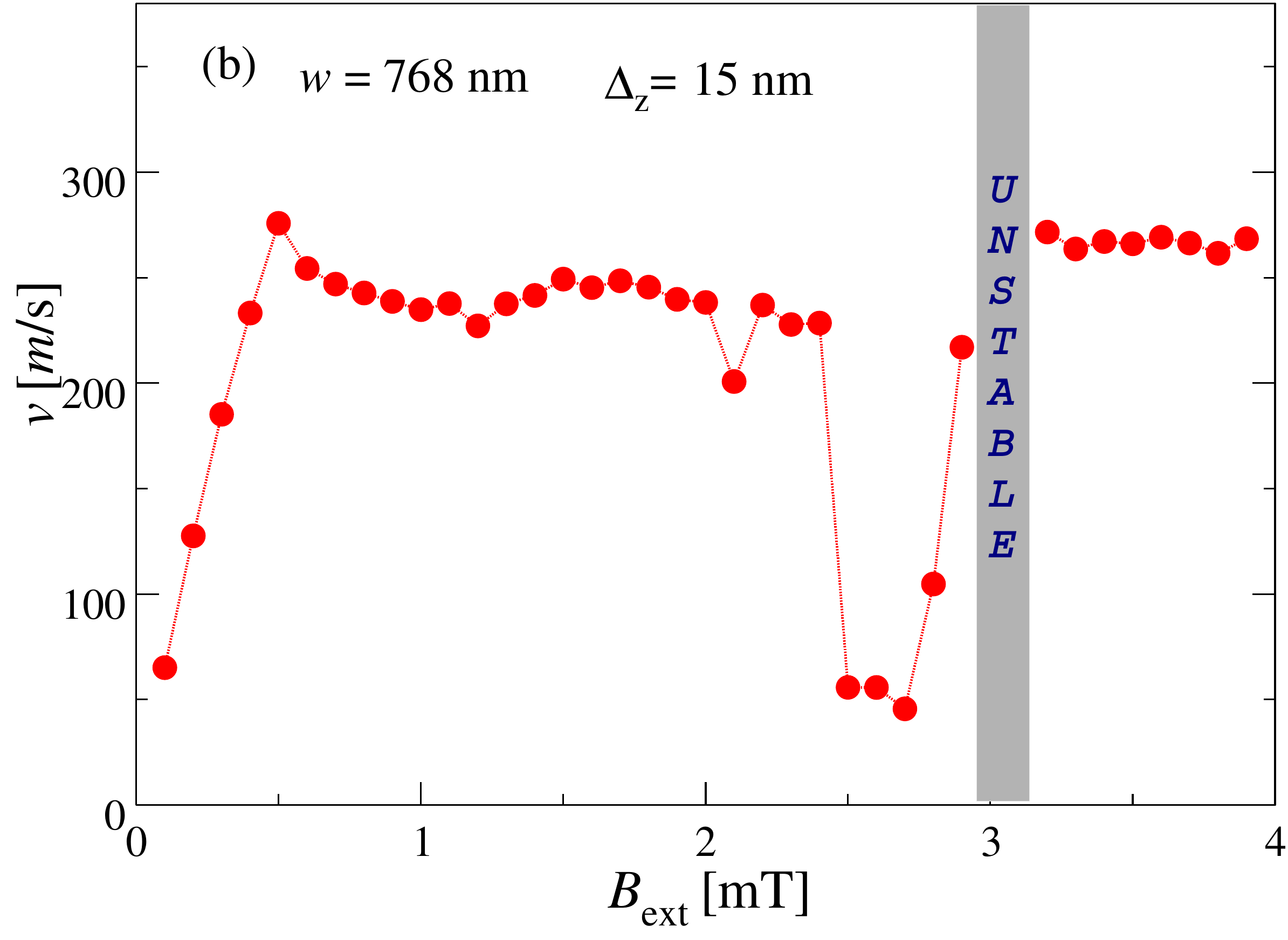}
\caption{(color online) $v(B_\text{ext})$ curves for different strip geometries, where 
the  VW is the equilibrium DW structure. (a) For two strips, both having the same
width of $w = 240$ nm and in-plane cell size 5 nm, and different thicknesses $\Delta_z = 10$ nm and 
$\Delta_z = 20$ nm. (b) The same for a strip with $w = 768$ nm, $\Delta_z = 15$ 
nm and in-plane cell size 3 nm. For 3 mT $\leq B_\text{ext} \leq$ 3.1 mT the DW is unstable; the DW width
grows without bound.}
\label{fig:fig2_dynamics}
\end{figure}

\section{VORTEX WALL DYNAMICS}
\label{vortex}

\begin{figure}[t!]
\leavevmode
\includegraphics[trim=0cm 0cm 0cm 0.5cm,clip=true,width=0.5\textwidth]{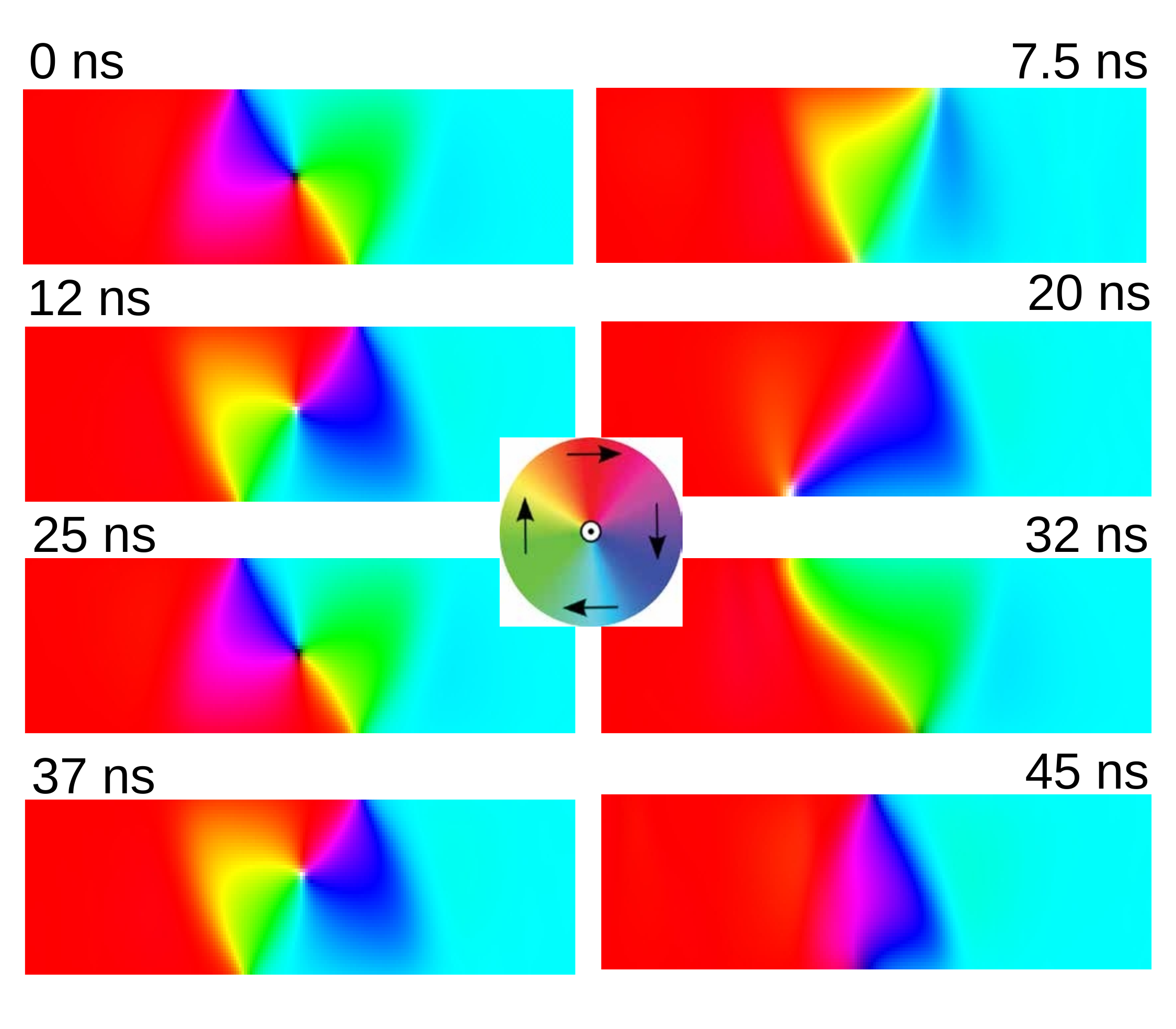}
\caption{(color online) Vortex wall dynamics in a strip with $w = 240$ nm 
and $\Delta_z = 10$ nm for $B_\text{ext} = 1.5$ 
mT (i.e. above the Walker field $B_\text{W} =1.2$ mT). 
The DW dynamics repeats a cycle where the vortex core first moves out of the strip through
the top strip edge, forming an ATW, followed by the injection of a vortex core from the top edge 
with the opposite polarity, which subsequently moves across the strip to the bottom edge, 
reversing the ATW magnetization. Then a vortex, again with a polarity opposite to that of
the previous one, is nucleated from the bottom edge, and moves to the top edge, again
reversing the ATW magnetization. Then the same process repeats.}
\label{fig:fig3_dynamics}
\end{figure}
 
The vortex wall is a DW structure characterized by the chirality or sense of rotation of the vortex, and the core polarity $p$. The latter may assume two different values, $p=\pm$1, corresponding to the two possible out-of-plane magnetization directions of the core. Vortex wall is the equilibrium DW structure within a wide range of $w$ and 
$\Delta_z$\cite{nosotrosequilibrium}, and thus it is a pertinent question to 
what extent the dynamics exhibited by it depends on the geometry of the strip.
To this end, in what follows, we consider some example cases, demonstrating
that already in relatively narrow strips, several different kinds of field-driven 
VW dynamics may be observed.

\subsection{Vortex wall dynamics in narrow strips}

To illustrate this, we start by considering the DW velocity $v$ as a function of
the applied field $B_\text{ext}$. 
 In Fig. ~\ref{fig:fig2_dynamics} we show that quite different $v(B_\text{ext})$ curves are obtained depending on $w$ and $\Delta_z$. Fig.~\ref{fig:fig2_dynamics} (a) 
shows that for a relatively narrow strip width $w$ = 240 nm, considering two different 
thicknesses ($\Delta_z =$ 10 nm and $\Delta_z =$ 20 nm) leads to clear differences 
in the $v(B_\text{ext})$ curves already below the $\Delta_z$ dependent Walker field. 
In the case of $\Delta_z=$ 10 nm, for very small fields $v$ increases linearly with $B_\text{ext}$, a behavior 
arising from steady VW motion with the vortex core assuming an off-centre position 
within the strip\cite{reviewthiaville}. For $B_\text{ext} \geq$ 0.4 mT the vortex 
core is expelled out of the strip due to the gyrotropic force\cite{malozemoff1979}, 
leading to a TW, and to a sub-linear $v(B_\text{ext})$ relation \cite{reviewthiaville}. In contrast, 
for the thicker strip with $\Delta_z =$ 20 nm, the linear dependence of $v$ on 
$B_\text{ext}$, originating from steady VW dynamics with the vortex core within the 
strip, persists up to the Walker field. Notice also that for $B_\text{ext}$ such 
that the VW has transformed into a TW for $\Delta_z =$ 10 nm, the TW velocity for 
a given $B_\text{ext}$ exceeds the corresponding VW velocity in the system with 
$\Delta_z =$ 20 nm. This is likely due to the large energy dissipation associated 
with the dynamics of the vortex core \cite{malozemoff1979}.

\begin{figure}[t!]
\leavevmode
\includegraphics[trim=0cm 0cm 0cm 0.5cm,clip=true,width=0.5\textwidth]{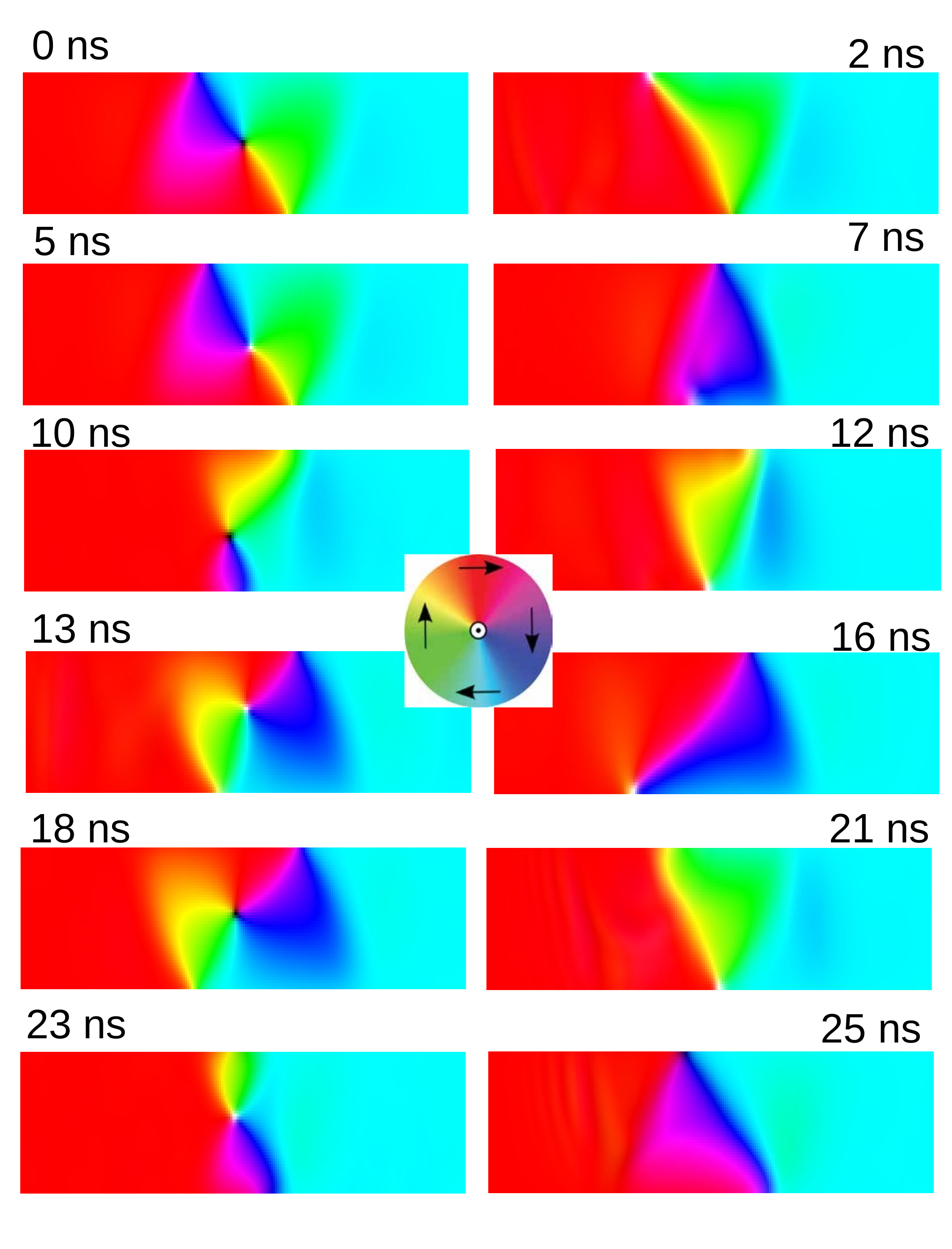}
\caption{(color online) Vortex wall dynamics in a strip with $w=240$ nm and 
$\Delta_z=10$ nm for  $B_\text{ext}=4$ mT (above the Walker breakdown, $B_\text{W}=1.2$ 
mT). Now the dynamics alternates between ATWs of different magnetization, 
with its reversal mediated by either a vortex or an antivortex core moving across 
the strip width.}
\label{fig:w240_V_t10_B4_dynamics}
\end{figure}

Above the Walker breakdown, $v$ drops dramatically as a consequence of  
the onset of periodic transformations between different types of DW structures. 
An example of this behavior is shown in Fig.~\ref{fig:fig3_dynamics}, where the VW 
dynamics for $B_\text{ext}=$ 1.5 mT, exceeding the Walker field $B_\text{W}$, is 
shown for a strip with $w =$ 240 nm and $\Delta_z$ = 10 nm (see also Supplemental Material Movie 1 \cite{SM}). Periodic 
transformations between VW and ATW structures are observed, mediated by the
transverse motion of the vortex core across the strip width. Notice that here
the vortex core dynamics alternates between injection from the top and bottom 
edges, followed by its top-to-bottom and bottom-to-top propagation across the 
strip, respectively. In Ref.\cite{reviewthiaville}, the same geometry and $B_\text{ext}$ were 
found to always lead to vortex core injection from the top strip edge, followed 
by its ``downwards'' motion towards the bottom edge. We have checked that this 
is due to slightly different values of $M_\text{s}$ and $A_\text{ex}$ 
(i.e. $M_\text{s}$ = $8 \times 10^{5}$ A/m and $A_\text{ex} = 10^{-11}$ J/m,
respectively) used in the simulations of Ref.\cite{reviewthiaville}. 
Given such sensitivity to small differences in the simulation conditions,
one may expect that also other kinds of VW dynamics may be observed.

An example thereof is provided in Fig.~\ref{fig:w240_V_t10_B4_dynamics}, where
the same geometry as above (i.e. $w =$ 240 nm and $\Delta_z$ = 10 nm) is
considered for a larger driving field of $B_\text{ext} = 4$ mT. Now, the
DW exhibits transformations between VW, ATW and AVW structures, with the
repeated ATW magnetization reversal mediated by alternating nucleation and 
propagation across the strip of either a vortex or an antivortex core (see also Supplemental Material Movie 2 \cite{SM}).
Thicker strips appear to lead to a smaller number of different periodicities.
For instance, for $w=240$ nm and $\Delta_z=20$ nm, only two periodicities are
observed for $B_\text{ext}>B_\text{W}$, consisting of transformations between
VW and ATW structures. While the ATW reverses its magnetization, the VW  structures exhibit alternating polarities but no change of chirality takes place (not shown). For some fields, also a nucleation of an antivortex in the edge 
of the strip is observed, followed by an vortex-antivortex annihilation process. Thus, already 
for relatively confined geometries, the field driven VW dynamics is quite complex,
given that different geometries and applied fields may lead to numerous different
dynamical behaviors. 

\subsection{Vortex wall dynamics in wide strips}

In the case of wider strips, the VW dynamics is even more complex. 
Fig.~\ref{fig:fig2_dynamics}(b) shows  the $v(B_\text{ext})$ curve for a strip 
with $w = 768$ nm and $\Delta_z = 15$ nm. As can be seen, the resulting 
$v(B_\text{ext})$ curve is very different to those observed for more narrow strips 
[Fig.~\ref{fig:fig2_dynamics} (a)], and consists of six different regimes. 
Analogously to more narrow strips, at low fields the velocity increases linearly 
with $B_\text{ext}$, with the vortex core assuming again a steady, $B_\text{ext}$ 
dependent off-centre position.

\begin{figure}[t!]
\leavevmode
\includegraphics[clip,width=0.45\textwidth]{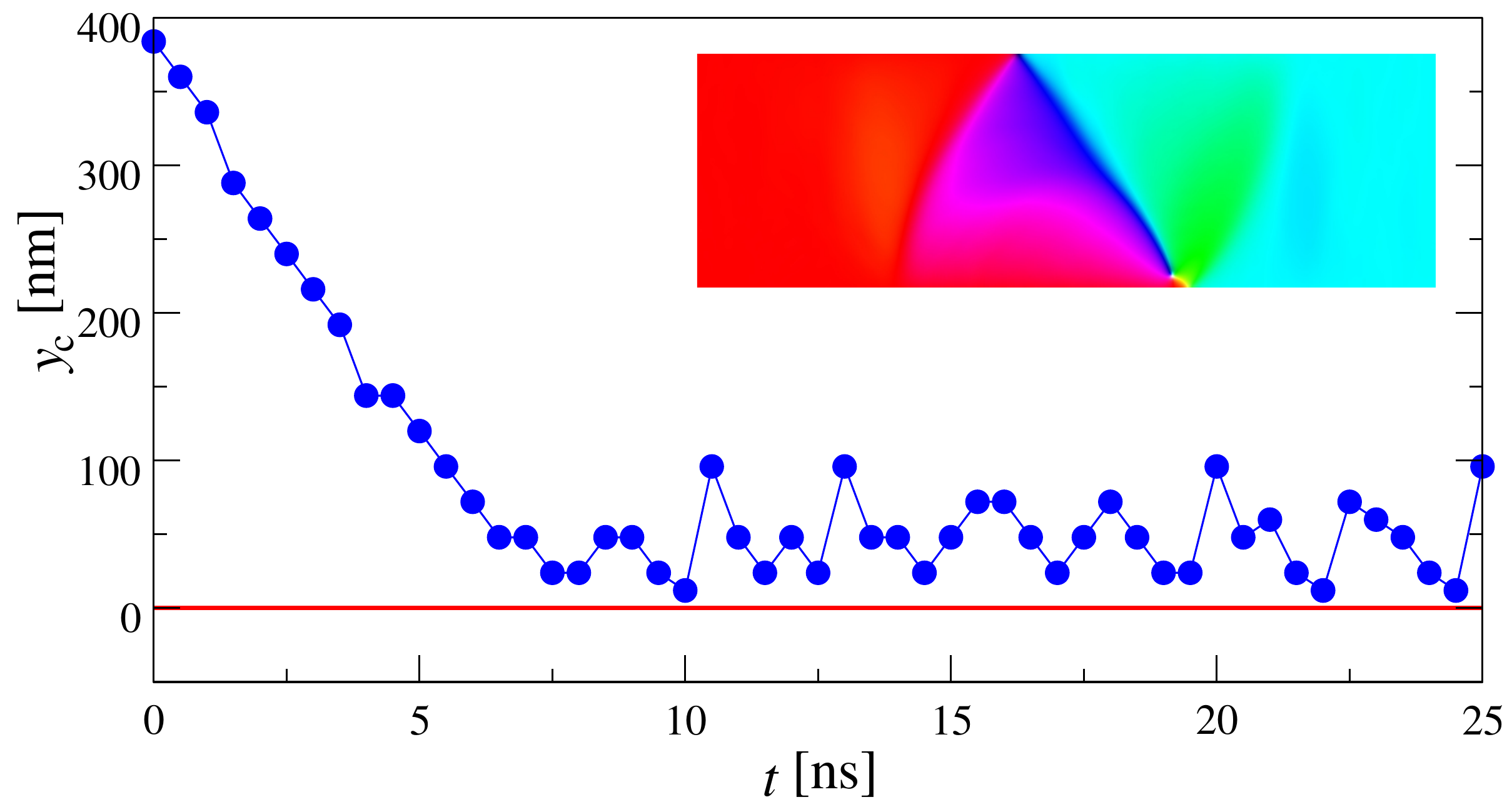}
\caption{(color online) The trajectory $y_\text c$ of the vortex core in the vicinity
of the edge of the system for $w=768$ nm, $\Delta_z= 15$ nm and $B_\text{ext}= 1.5$ mT,
illustrating the attraction-repulsion effect. The red line at $y_c=$ 0 corresponds to the strip edge. The inset shows an example
of the DW configuration with the vortex core close to the edge of the strip.}
\label{fig:fig3_2_dynamics}
\end{figure}

For 0.6 mT $\leq B_\text{ext} \leq 2.4$ mT, the $v(B_\text{ext})$ curve initially
exhibits a small, negative gradient, but looks otherwise essentially like
a plateau. Thus, the DW is able to maintain a relatively high velocity of
approximately 250 m/s, while for more narrow strips [Fig.~\ref{fig:fig2_dynamics}(a)]
Walker breakdown leads to significantly smaller DW velocities for most of this
range of $B_\text{ext}$. This behavior arises as a consequence of the vortex
core moving towards the strip edge due to the gyrotropic force, but contrary
to the behavior observed in more narrow strips, it is not able to leave the strip.
Instead, the competition between the gyrotropic force and a repulsive interaction of 
the vortex core with the half-antivortex (HAV) edge defect \cite{ZIN-11} leads to 
periodic oscillations of the vortex core $y$-position in the vicinity of the 
strip edge. Fig.~\ref{fig:fig3_2_dynamics} shows an example of the trajectory 
($y$ coordinate $y_\text{c}$ as a function of time $t$) of the vortex core in a strip 
with $w = 768$ nm and $\Delta_z = 15$ nm with the DW subject to an applied 
field of $B_\text{ext} = 1.5$ mT. A snapshot of the DW configuration with
the vortex core close to the HAV edge defect is included. An example of this process is 
also shown in the Supplemental Material Movie 3 \cite{SM}. The apparent emissions of spin waves observed in the movie may be related to short-lived (faster than the frame rate) transient dynamics involving nucleation/annihilation processes. Such behavior arises in wide strips due to the interplay between the gyrotropic force and the interaction of the vortex core with the HAV being different 
to the one in narrow strips. Since 
in permalloy everything is dominated by shape anisotropy originating from the 
strip edges, the energy needed to displace the vortex core from the middle of 
the strip towards the edge by the gyrotropic force is smaller than in narrow 
strips. This is evidenced also by the steep increase of $v(B_\text{ext})$ in 
Fig.~\ref{fig:fig2_dynamics}(b) for small $B_\text{ext}$. At the same time, 
the HAV edge defect is more localized in relative terms in wider strips, 
thus leading to a relatively strong short-range vortex core-HAV interaction, 
preventing the former from being expelled from the strip. In other words, the 
energy necessary to move the vortex core from its equilibrium position in 
the middle of the strip towards the strip edge is smaller than the energy needed 
to push the core out of the strip. As a result, the core exhibits oscillatory
dynamics close to the strip edge, and the DW assumes a relatively high
propagation velocity. In what follows, we shall refer to this phenomenon 
as the attraction-repulsion effect.

\begin{figure}[t!]
\leavevmode
\includegraphics[clip,width=0.49\textwidth]{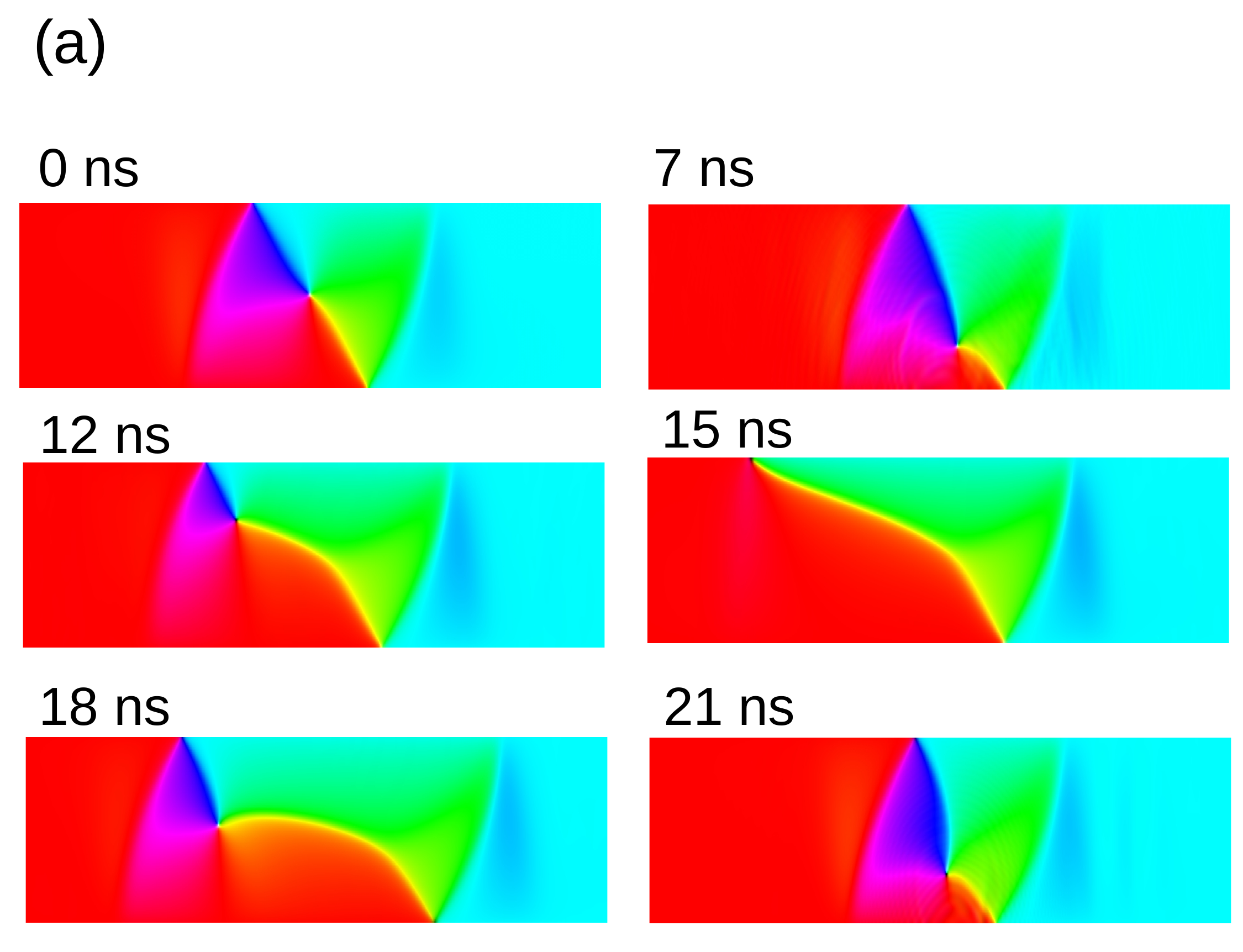}
\includegraphics[trim=0.5cm 0cm 0.5cm 0cm,clip=true,width=0.49\textwidth]{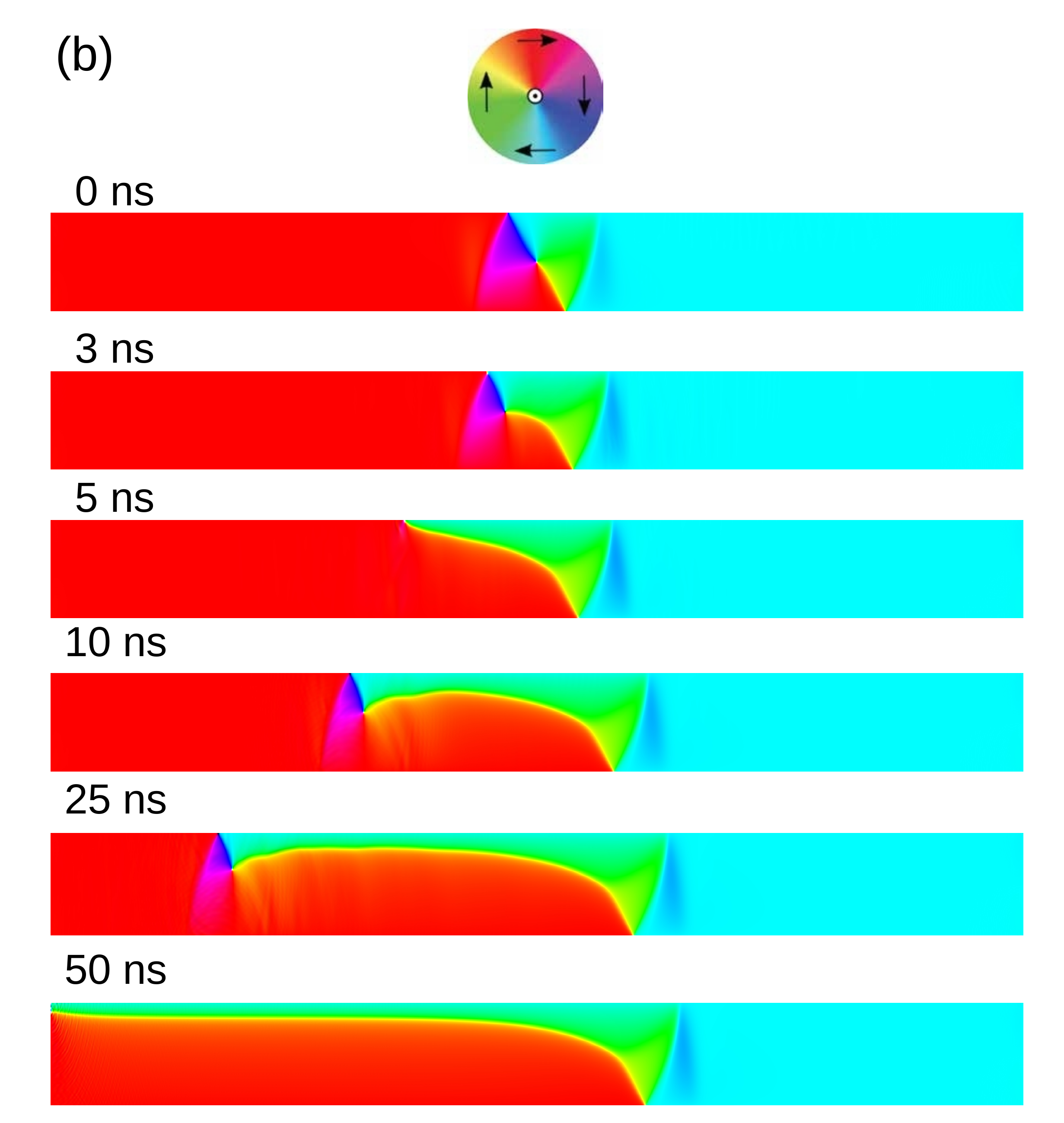}
\caption{(color online)(a) Walker breakdown-type of dynamics for 
$w=768$ nm, $\Delta_z = 15$ nm and $B_\text{ext}=$ 2.5 mT. The vortex
core exhibits asymmetric lateral oscillations, leaving the strip at the
top edge, but reversing the direction of motion and polarity of the core 
well inside the strip when closer to the bottom edge.
(b) The instability occuring for $B_\text{ext}=3$ mT, where the DW width
grows without limit as the leading HAV edge defect moves faster than the
combination of the trailing HAV and the vortex core.}
\label{fig:fig4_dynamics}
\end{figure}

This behavior is characteristic of all field values within the range 
0.6 mT $\leq B_\text{ext} \leq 2.4$ mT, with the exception of 
$B_\text{ext}=2.1$ mT. Here, $v$ is a little smaller than elsewhere in 
the plateau region visible in Fig.~\ref{fig:fig2_dynamics} (b). 
For that case, although most of the time the usual attraction-repulsion
effect takes place, in rare instances the repulsion effect is strong enough to 
displace the vortex core all the way to the other edge of the strip, leading to a structural DW transformation (VW to ATW), and 
to a reduced DW propagation velocity. 

For 2.5 mT $\leq B_\text{ext} \leq 2.7$ mT, $v$ decreases abruptly 
[Fig.~\ref{fig:fig2_dynamics} (b)]. In this regime, the DW dynamics
exhibits repeated transitions between a VW and a kind of stretched TW
(or ATW) structure, see Fig. \ref{fig:fig4_dynamics}(a). The peculiar feature
of this Walker breakdown -like dynamics is that while the vortex core is 
able to reach the top edge of the strip, transforming the DW momentarily 
into a TW-like structure, the vortex core never reaches the opposite (bottom) 
edge of the strip. Thus, the transient TW-like structure always has the 
same magnetization. The resulting oscillatory dynamics of the vortex core resembles 
that observed above in the context of the attraction-repulsion effect. However,
the amplitude of the transverse core oscillations is significantly larger,
leading to a higher rate of energy dissipation associated with the internal
dynamics of the DW, resulting also in a significantly reduced DW propagation
velocity.

For $B_\text{ext}=2.8$ mT and $B_\text{ext}=2.9$ mT, the velocity increases as the 
attraction-repulsion effect appears again. In the case of $B_\text{ext}=2.8$ mT although
also an attraction-repulsion effect is observed, the transformations between different DW structures (VW and ATW) are predominant. For $B_\text{ext}=2.9$ mT 
only the attraction-repulsion effect is observed, leading to a higher DW propagation
velocity [Fig.~\ref{fig:fig2_dynamics} (b)]. 

For $B_\text{ext} = 3$ mT and $B_\text{ext} = 3.1$ mT, the leading and trailing
edges of the DW start to move at different velocities, implying that the DW width grows
without limit, and thus the system is unable to support a compact DW structure.
The leading HAV moves faster than the trailing one, and the vortex core exhibits
oscillatory transverse motion in the vicinity of the latter, see 
Fig. \ref{fig:fig4_dynamics} (b). Eventually the trailing edge of the DW exits the
computational window.  Such unstable DW dynamics has previously been
reported in Ref.\cite{ZIN-11}, and renders the measurement of a well-defined DW
velocity impossible. 

\begin{figure}[t!]
\leavevmode
\includegraphics[clip,width=0.5\textwidth]{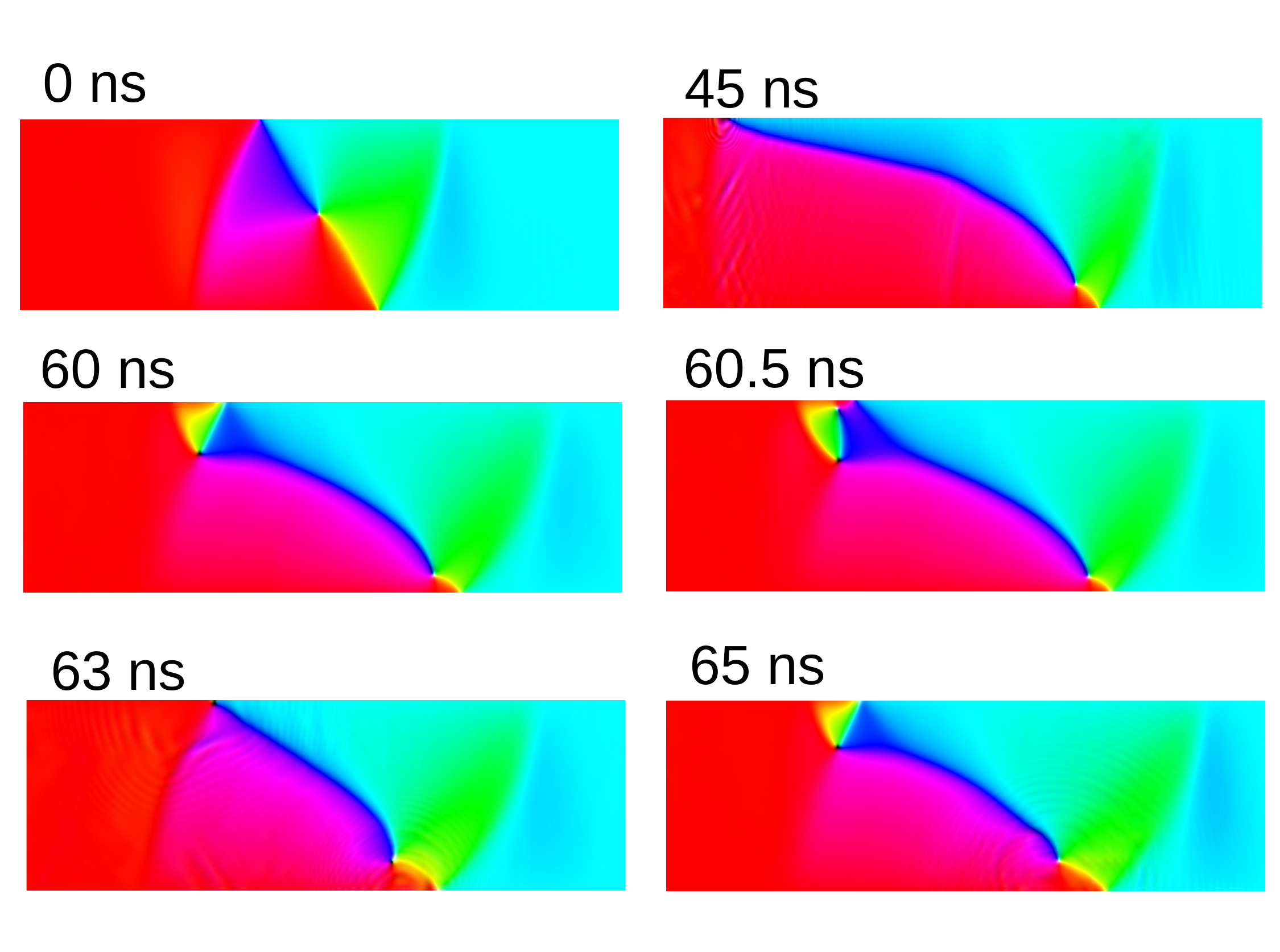}
\caption{(color online)  VW dynamics for $B_\text{ext}=3.3$ mT in a strip 
with $w = 768$ nm and $\Delta_z = 15$ nm. The trailing HAV edge defect 
repeatedly emits a vortex-antivortex pair, which then annihilates, and the 
process is repeated.}
\label{fig:fig5_dynamics}
\end{figure}

Curiously, for larger fields (3.2 mT $\leqslant B_\text{ext}\leqslant 4$ mT;
we focus here on fields up to 4 mT) the DW structure is stable again, and 
the $v(B_\text{ext})$ curve displays a second high-velocity plateau with $v$ 
exceeding 250 m/s. There, after a relatively long initial transient, the
DW dynamics proceeds as follows: In the vicinity of the leading edge HAV, the vortex
core exhibits similar oscillatory dynamics as in the attraction-repulsion
effect. The trailing HAV repeatedly emits a vortex-antivortex pair, which
subsequently annihilates. Fig. \ref{fig:fig5_dynamics} and the Supplemental
Material Movie 4 \cite{SM} illustrate this dynamics. Notice that while the above-mentioned
sequence is the dominating one, sometimes the vortex-antivortex pair fails 
to annihilate in the vicinity of the trailing HAV, and is able to move to 
the opposite edge, leading to a short period of more complex dynamics before
the above-mentioned sequence starts again (see Supplemental Material
Movie 4 \cite{SM}). Similar vortex-antivortex annihilation processes within DWs have 
previously been reported in strips with much more confined lateral 
dimensions, and applied fields roughly 5 times as high as here\cite{kim2008}.

Vortex-antivortex annihilation processes can be parallel or antiparallel depending on the polarity of the vortex-antivortex pair\cite{hertel2006,waeyenberge2006,tetriakov2007}. When the vortex and the antivortex have the same polarity (parallel case), the process is continuous, whereas for the antiparallel one there is emission of spin waves as a result of the annihilation. This is related to the topological charges (winding number $n$) and the polarities of the defects. DW structures are composed of topological defects, which have associated winding numbers $n=$ +1 for vortices, -1 for antivortices, and $\pm$ 1/2 for edge defects\cite{TCH-05}. In DW structures the topological defects are compensated, leading to zero total winding number, see Fig. \ref{fig:topological}. However, there are also other ``topological charges'' associated with the winding number and the polarity of the core of vortices and antivortices, e.g. the skyrmion charge $q=np/2$\cite{belavin1975}. For parallel polarity, vortex and antivortex have opposite skyrmion numbers, leading to a continuous annihilation process. However, for the antiparallel case the skyrmion numbers are equal, and there is no compensation. As a result,  emission of spin waves occurs after the annihilation process\cite{hertel2006,waeyenberge2006,tetriakov2007}. Here,  we have  observed both parallel and antiparallel  processes of annihilation of the vortex-antivortex pair.

\begin{figure}[t!]
\leavevmode
\includegraphics[clip,width=0.5\textwidth]{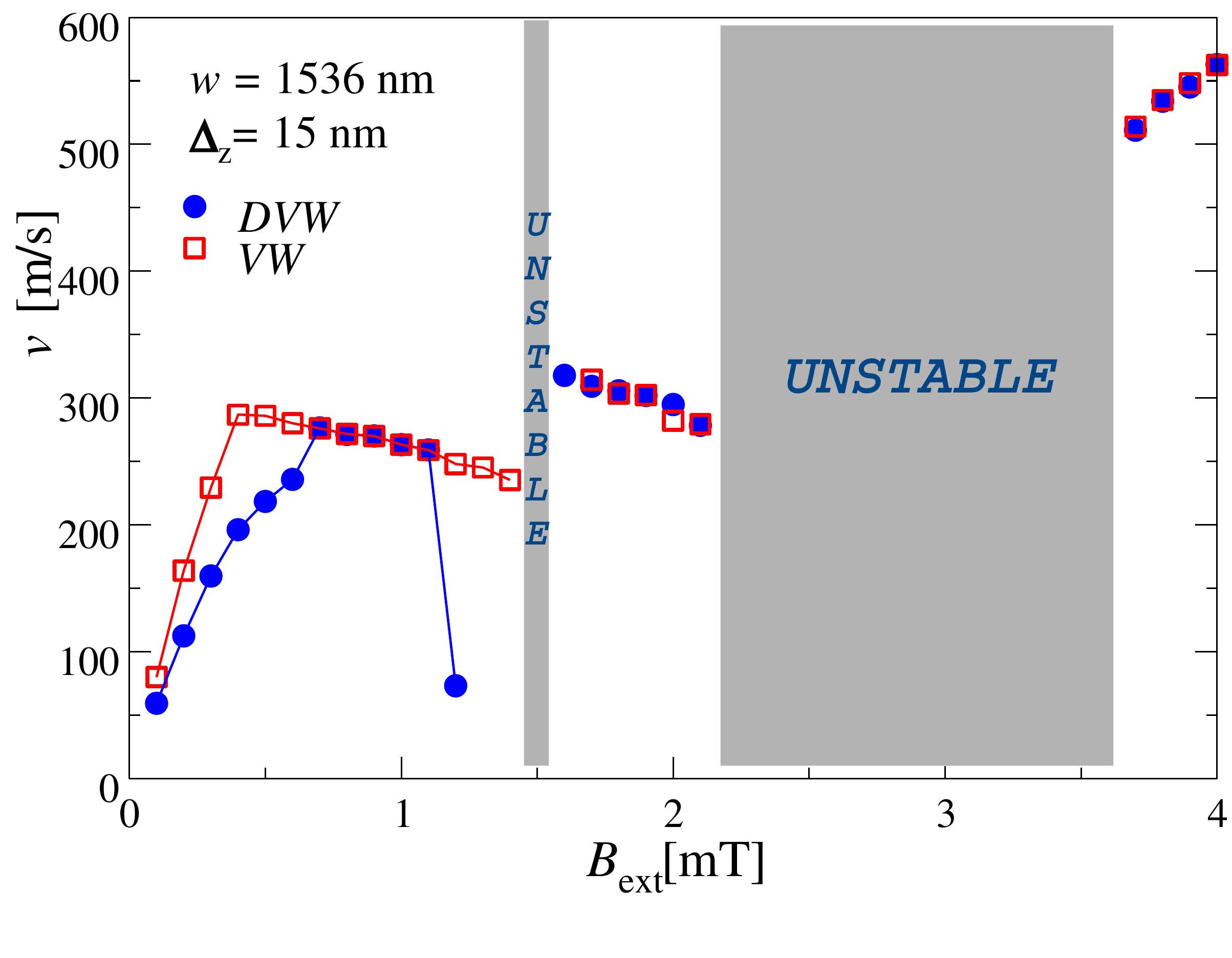}
\caption{(color online) $v(B_\text{ext})$ curves for a VW and a DVW initial state 
in a strip with $w=1536$ nm and $\Delta_z = 15$ nm, where both structures are 
equilibrium DWs. In-plane cell size equal to 3 nm. The gray areas indicate unstable regions (no compact DWs).}
\label{fig:fig7_dynamics}
\end{figure}

For even larger strips the VW dynamics is again found to be different from
that observed in more narrow systems. For example, let us consider a strip 
with $w = 1536$ nm and $\Delta_z = 15$ nm, for which both VW and DVW are
equilibrium structures (i.e. they have the same energy \cite{nosotrosequilibrium}).
The $v(B_\text{ext})$ curve obtained when starting from a VW initial state
is shown as red squares in Fig. \ref{fig:fig7_dynamics}. As before, the low
field steady/viscous regime (for $B_\text{ext} \leq$ 0.4 mT) is followed
by a plateau-like regime where the velocity decreases slowly with the field 
(for 0.5 mT $\leq B_\text{ext} \leq$ 1.4 mT), due to the attraction-repulsion 
effect. Contrary to the case of the more narrow strip considered in 
Fig.~\ref{fig:fig2_dynamics} (b), here no Walker breakdown is observed. Instead, 
the attraction-repulsion regime is terminated by an instability where the 
DW width grows without bound for $B_\text{ext}=1.5$ mT and $B_\text{ext}=1.6$ mT.

\begin{figure}[t!]
\leavevmode
\includegraphics[trim=0cm 0cm 0cm 0.5cm,clip=true,width=0.5\textwidth]{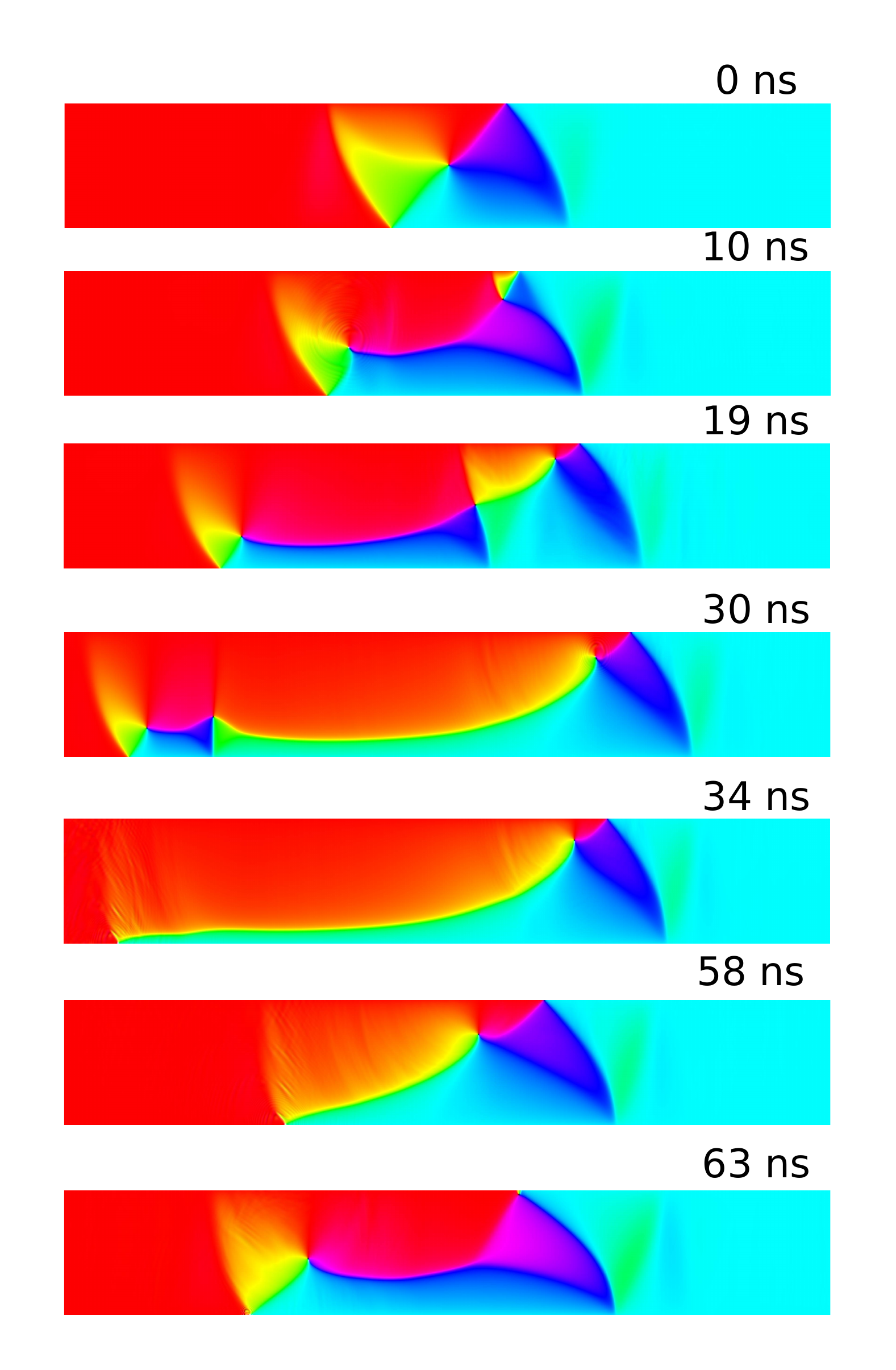}
\caption{(color online) DW dynamics starting from a VW initial state for 
 $w=1536$ nm, $\Delta_z=15$ nm and $B_\text{ext}=2.1$ mT. The DW dynamics shows the nucleation of an antivortex into the strip from the top edge, followed by the injection of a vortex. The antivortex moves towards the vortex that was previously in the strip, leading to an annihilation process between them. After that, the DW structure reduces its width, resulting a VW. Then the process is repeated.}
\label{fig:fig9_dynamics}
\end{figure}

Above this unstable regime, the velocity again                                  
decreases linearly with the field (1.7 mT $\leq B_\text{ext} \leq$ 2.1 mT). 
An example of the type of DW dynamics observed in this regime is shown in Fig.~\ref{fig:fig9_dynamics} for $B_\text{ext}=2.1$ mT (see also Supplemental Material Movie 5 \cite{SM}). When the field is applied the vortex core is displaced from its equilibrium position, and an antivortex enters into the strip. After that, a vortex nucleates into the strip. The antivortex moves towards the previous vortex, and an annihilation process between them takes places (snapshot at 34 ns shows the DW structure just after the annihilation). Unlike in the case observed for the strip with $w=768$nm and $\Delta_z=15$nm, where the  processes of annihilation were between the vortex and antivortex that nucleate into the strip, now the annihilation is between the antivortex and the vortex that was inside the strip previously. Once the annihilation process occurs, this part of the DW structure starts to move towards the other vortex, reducing the width of the DW structure. Again two parts of the DW structure are moving with different velocities. The vortex that does not participate in the annihilation is feeling  an attraction-repulsion effect. Although this is the main dynamical sequence, sometimes other more complex processes are observed, such as one involving a vortex and two antivortices. 

For $2.2 \leq B_\text{ext} \leq$ 3.6 mT, an extended region where the
system is again unable to support compact DWs is encountered. Two different types of instability have been observed. For $2.2 \leq B_\text{ext} \leq$ 3.4 mT the instability is like the one shown in  Fig.~\ref{fig:fig4_dynamics} (b), where the DW width grows without limit. For  $B_\text{ext}=$ 3.5 mT and $B_\text{ext}=$ 3.6mT, although the DW width also increases without limit, at the same time several vortex-antivortex annihilation processes take place. 

\begin{figure}[t!]
\leavevmode
\includegraphics[trim=0cm 0cm 0cm 1cm,clip=true,width=0.5\textwidth]{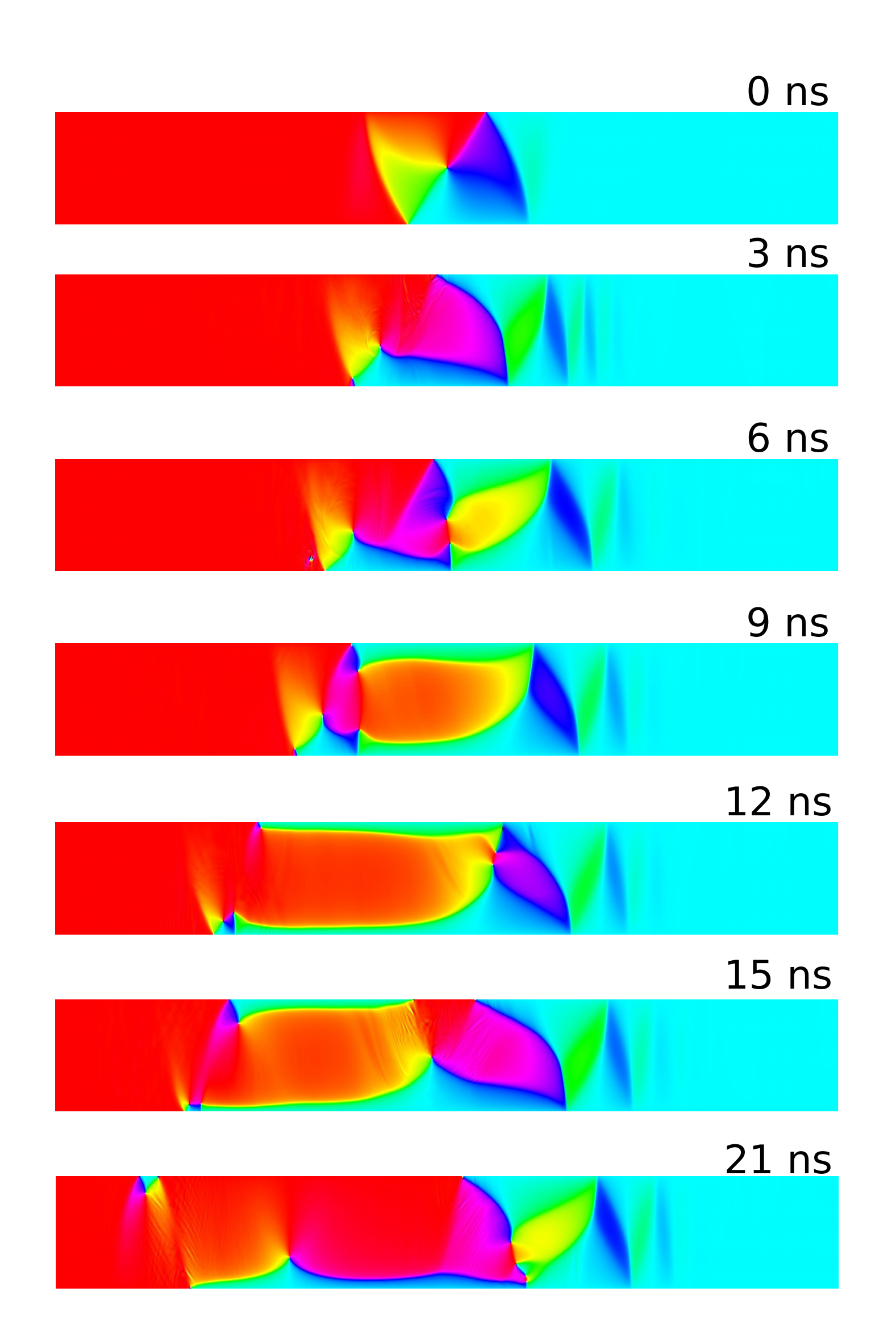}
\caption{(color online) 
DW dynamics starting from a VW initial state for $w=1536$ nm, 
$\Delta_z=15$ nm and $B_\text{ext}=3.8$ mT. Very complex, possibly chaotic 
dynamics is observed.}
\label{fig:fig10_dynamics}
\end{figure}

For even larger fields,
compact DWs are observed again, with their large velocity increasing linearly
with $B_\text{ext}$. Notice that for $B_\text{ext}=4$ mT, $v$ exceeds 550 m/s. In this
regime, the DW dynamics is quite chaotic with transformations between  
complex dynamical states. In Fig.~\ref{fig:fig10_dynamics} 
the DW dynamics is shown for $B_\text{ext}=3.8$ mT (see also Supplemental Material 
Movie 6\cite{SM}). Repeated annihilation processes between vortices and 
antivortices are observed. Note that in this regime, creation of vortices 
and antivortices occurs not only at the edges (as for lower fields), but 
also within the strip (see  snapshots at 6, 12 and 21 ns in Fig.~\ref{fig:fig10_dynamics}).
Similar processes of vortex-antivortex pair generation within the strip
have been reported before \cite{zeisberger2012}.

\begin{figure}[t!]
\leavevmode
\includegraphics[clip,width=0.45\textwidth]{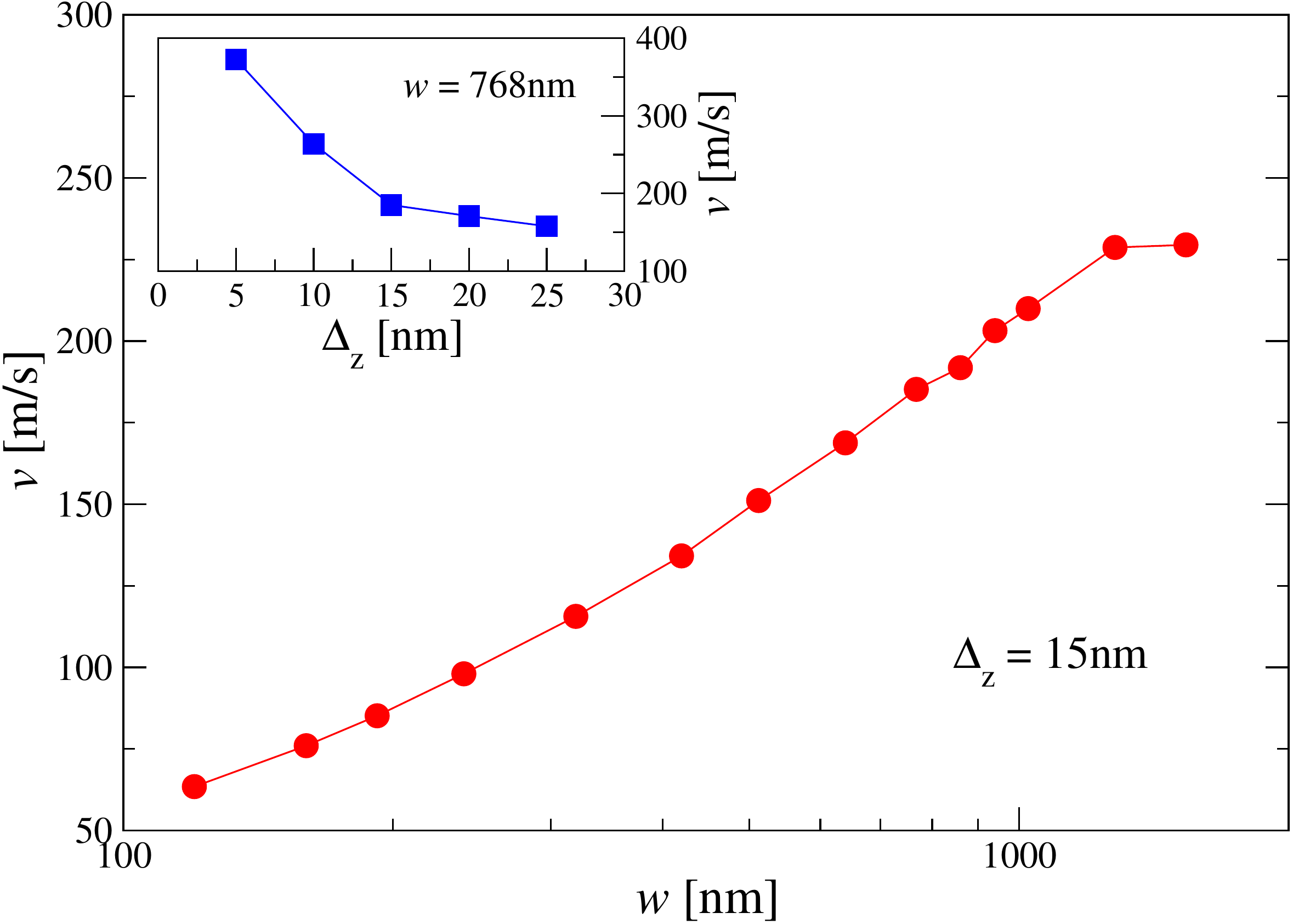}
\caption{(color online) Main figure: VW velocity $v$ as a function of the 
width of the strip $w$ when the thickness and field are fixed to $\Delta_z=15$ nm 
and $B_{ext}=0.3$ mT, respectively. $w$ is in logarithmic scale. Inset: VW velocity $v$ as a function of $\Delta_z$ 
for $w=768$ nm and $B_\text{ext}=0.3$ mT.}
\label{fig:fig6_dynamics}
\end{figure}

We have also analyzed the dependence of the VW velocity within the low-field
steady/viscous regime as a function of $w$ and $\Delta_z$,
within a range of $w$ and $\Delta_z$ such that VW is the equilibrium structure. 
The main panel of Fig.~\ref{fig:fig6_dynamics} shows the VW velocity as a 
function of $w$ for $\Delta_z=15$ nm and $B_\text{ext}=0.3$ mT, displaying
an increasing trend of $v$ with $w$. Since the width of VWs in wider strips 
is larger, such a trend is in agreement with the 1d model, predicting a linear
dependence of $v$ on the DW width \cite{reviewthiaville}. This can also be
rationalized by the fact that a given amount of spin rotation leads to
DW displacements proportional to the DW width. Inset of Fig.~\ref{fig:fig6_dynamics} 
displays $v$ as a function of $\Delta_z$ for $w=768$ nm and $B_\text{ext}=0.3$ mT,
showing that the VW velocity decreases with the strip thickness. This may 
be attributed to the fact that in the steady state, the DW propagation occurs 
by a precessional motion around an out-of-plane demagnetizing or stray field, 
which is created by precession in the external field\cite{nakatani2005}. As the 
thickness of the strip increases, this demagnetizing field decreases, leading to 
a decreasing DW velocity.  

\section{DOUBLE VORTEX WALL DYNAMICS}
\label{doublevortex}

\begin{figure}[t!]
\leavevmode
\includegraphics[trim=1cm 0cm 1cm 0.5cm,clip=true,width=0.45\textwidth]{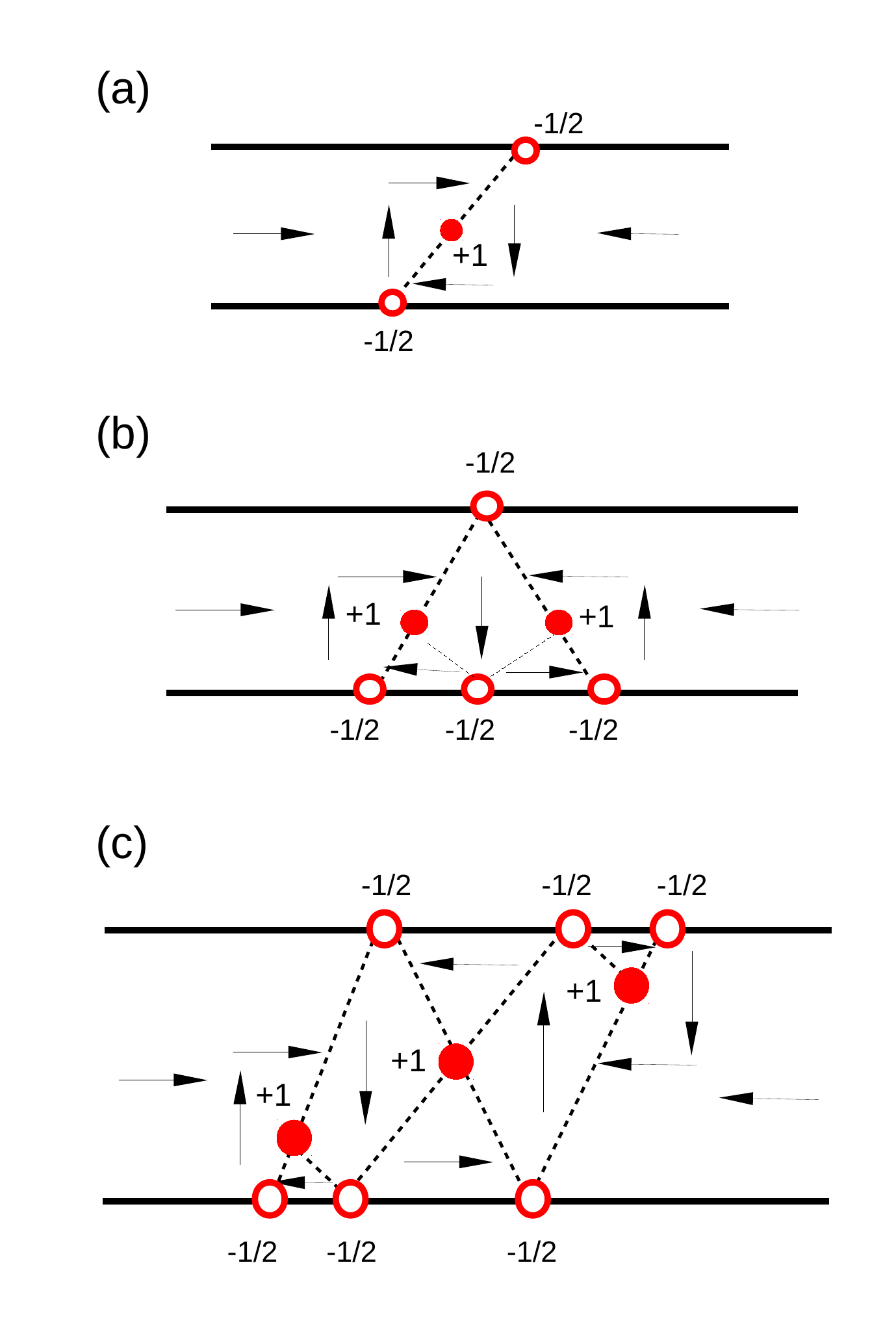}
\caption{(color online) The equilibrium DW structures in terms of their constituent 
elementary topological defects, i.e. vortices with a winding number +1 (full circles) and edge 
defects with winding number -1/2 (empty circles).
(a) Vortex wall, (b) double vortex wall, and (c) triple vortex wall.}
\label{fig:topological}
\end{figure}

For larger strips with widths around $1\mu$m and above, the DVW is the equilibrium 
DW structure\cite{nosotrosequilibrium}. This structure includes two 
vortices with opposite sense of rotation, see Fig.~\ref{fig:fig1_dynamics}(b). As occurs for the VW, the 
DVW presents different dynamical behaviors depending on the lateral dimensions 
of the strip, and the magnitude of the driving field. Moreover, in the case of  DVW the polarities of the two vortices are an important factor that strongly affects  the dynamics. For small enough $B_\text{ext}$ the DVW structure is stable depending on the polarities of the vortices. For the case in which the DVW is stable, the DW dynamics exhibits a unique low-field regime. However, the topology of the DVW, consisting of two vortices and 
4 HAV edge defects\cite{TCH-05} [see Fig. \ref{fig:topological}(b)] implies 
that the energy barrier for transforming the DVW into 
a VW is not very high. Thus, already for relatively low fields, DVWs tend to 
transform into VWs, via a process in which one of the vortices annihilates with
two edge defects at the edge of the strip. Since the total winding number of
a vortex and two edge defects is zero, this process happens easily, i.e. it does
not require a large applied field. 

\begin{figure}[t!]
\leavevmode
\includegraphics[trim=0cm 0cm 0cm 0.5cm,clip=true,width=0.5\textwidth]{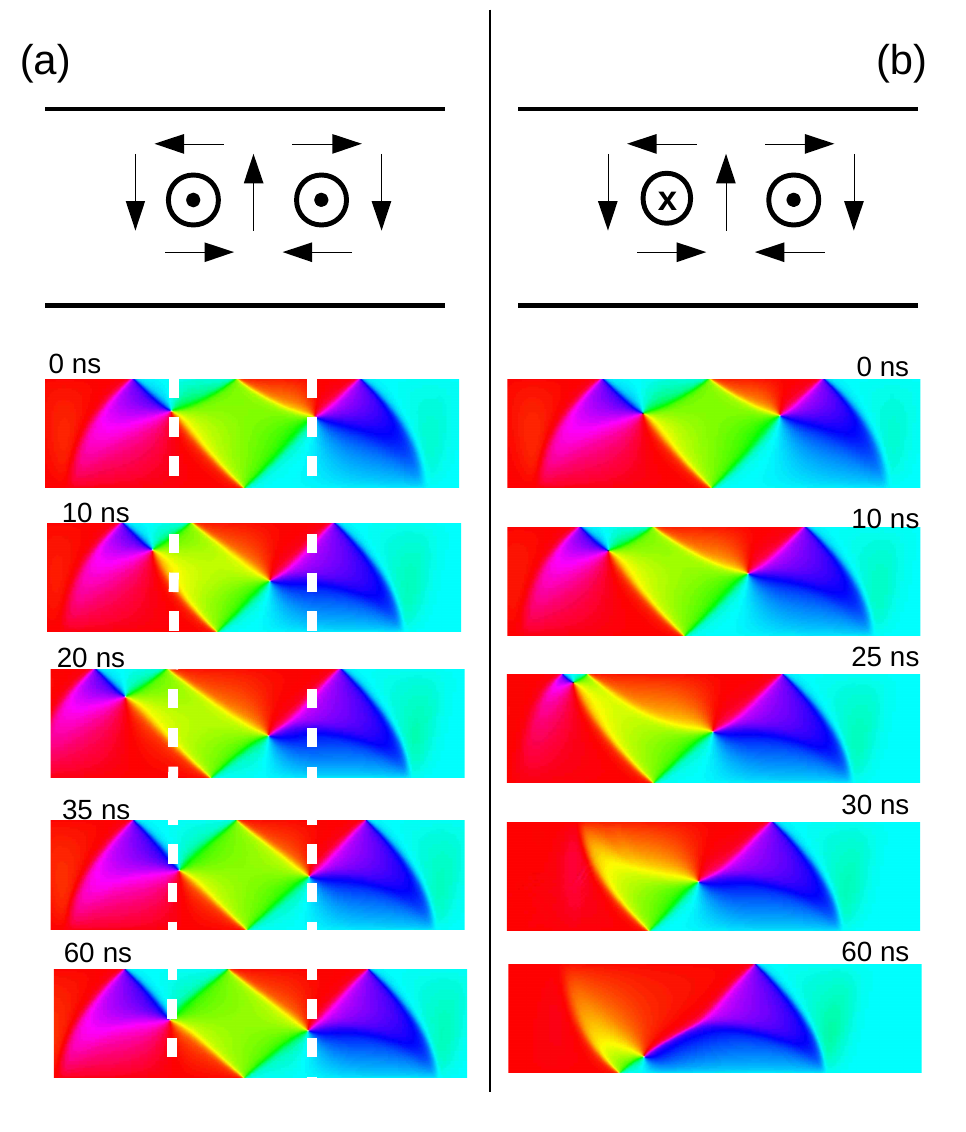}
\caption{(color online) DVW dynamics in a strip with $w=1536$ 
nm and $\Delta_z=15$ nm for $B_\text{ext}=0.3$ mT and different polarities of the vortices. 
(a) Parallel polarity; dashed lines help to see the transient relative motion of the vortex cores,
before the steady state motion with a non-evolving DVW structure is reached.
(b) Antiparallel polarity; DVW is not stable, and becomes a VW. }
\label{fig:fig8_dynamics}
\end{figure}

This can be illustrated by considering e.g. a strip with $w=1536$ nm and $\Delta_z=15$ 
nm, corresponding again to the phase boundary between VW and DVW equilibrium 
structures where they have the same energy \cite{nosotrosequilibrium}. The 
$v(B_\text{ext})$ curve obtained by starting from a DVW initial DW with equal polarity for both vortices (parallel case) is represented in 
Fig.~\ref{fig:fig7_dynamics}, together with the corresponding data for the VW
initial state in the same strip geometry (see the previous Section). For low 
fields ($B_\text{ext} \leqslant 0.6$ mT), the $v(B_\text{ext})$ curves are 
quite different for the two structures, indicating that the DVW structure is 
stable against small applied fields, and thus exhibits a unique small field 
steady/viscous regime. In this regime, the moving DVW structure initially 
exhibits transient oscillatory relative motion of the two vortex cores, with the
oscillation amplitude decreasing with time and eventually vanishing when the 
regime with steady DVW motion is reached [see Fig. \ref{fig:fig8_dynamics} (a), 
and Supplemental Material Movie 7\cite{SM}]. In this low-field regime the VW moves
faster than the DVW. This is due to the increased energy dissipation related to the
second vortex core of the DVW. 

However, if the polarities of the vortices are opposite (antiparallel case), the dynamics at low fields is very different. DVW is unstable and transforms into a VW, see Fig.\ref{fig:fig8_dynamics} (b). Contrary to the static case where the polarity barely contributes to the energy of the equilibrium state\cite{nosotrosequilibrium}, in the dynamics it plays an important role. This is because when an external field is applied, the vortex core feels a gyrotropic force\cite{malozemoff1979}  ${\bf F^{g}}=pG\bf \widehat{z}\times \bf v$, that depends on the polarity; $G=2\pi Jt$ is the gyrotropic constant and $\bf v$ the velocity. In the case of a VW  this force tries to expel
the vortex out through the top or bottom  edge depending on the polarity,  and independently of the vortex chirality. For DVWs with parallel polarity [see the panel top of Fig.\ref{fig:fig8_dynamics} (a)],  both vortices will try to leave the strip through the same edge due to an equal gyrotropic force exerted on them. As in a DW structure the topological defects have to be compensated for, i.e. the total winding number must be equal to zero, only one of the vortices can be expelled. As a result, a competition between them appears, which tends to keep both vortices in the strip, see Fig. \ref{fig:fig8_dynamics} (a). 

In the case of DVW with antiparallel polarity [see top in Fig.   \ref{fig:fig8_dynamics}(b)], the gyrotropic force on one of the vortices is opposite to the one exerted on the other. This means that one vortex will move to the top  edge, and the other to the bottom one. As a consequence one of the vortices will be expelled, i.e. the one located closer to the edge in its direction of movement. Fig. \ref{fig:fig8_dynamics} (b) shows how the DVW transforms into a VW.
 
Returning to the parallel case, for $B_\text{ext}$ = 0.7 mT, the external field is strong enough to
push one of the vortices out of the system (a process that takes place via annihilation
of the vortex with the two HAV edge defects), and the DW transforms into a VW.
Then, the steady state velocity equals that of the corresponding VW initial state, 
a behavior that persists up to $B_\text{ext}$ = 1.1 mT (see Fig. 
\ref{fig:fig7_dynamics}). Thus, also here the attraction-repulsion effect is
observed. However, the DW dynamics exhibits some dependence on the initial 
state also for fields exceeding 1 mT. For $B_\text{ext}$ = 1.2 mT, the DVW
initial state leads to a Walker breakdown-type of behavior, with repeated
transitions between different DW structures (VW and ATW), similarly to the observations in Fig.~\ref{fig:fig4_dynamics} (a). However, the corresponding system with a VW initial state displays the 
attraction-repulsion effect, and a significantly larger $v$ for the same field. Also, while 
for a VW initial state $B_\text{ext}$ = 1.3 mT and 1.4 mT lead to stable 
attraction-repulsion type of behavior, starting from a DVW initial state leads 
to unstable behavior where the DW width grows without limit. For $B_\text{ext}$ 
= 1.5 mT, such an instability is encountered also when starting from the 
VW initial configuration. For $B_\text{ext}$ =1.6 mT the instability occurs when the VW is the initial state, but not for the case in which DVW is the initial configuration. For larger fields, the two initial conditions lead 
to the same steady state dynamics, described in the previous section for the 
VW initial state, see Figs. \ref{fig:fig9_dynamics} and \ref{fig:fig10_dynamics}.  

Similar DVW dynamics is observed also for other $w$ and $\Delta_z$ values
with DVW as the equilibrium structure. However, in the parallel case the ``critical'' field
magnitude at which one of the two vortices is expelled from the strip,
leading to a VW, depends on $w$ and $\Delta_z$ in a non-trivial fashion.
For some systems belonging to the DVW equilibrium phase
this field is quite small, around 0.1 mT. A possible explanation is that 
the energy required to move one of the vortices towards the strip edge is
smaller in wider strips, given that shape anisotropy (an edge effect) is 
weaker in wider strips.

\section{TRIPLE VORTEX WALL DYNAMICS}
\label{triplevortex}

\begin{figure}[t!]
\leavevmode
\includegraphics[clip,width=0.5\textwidth]{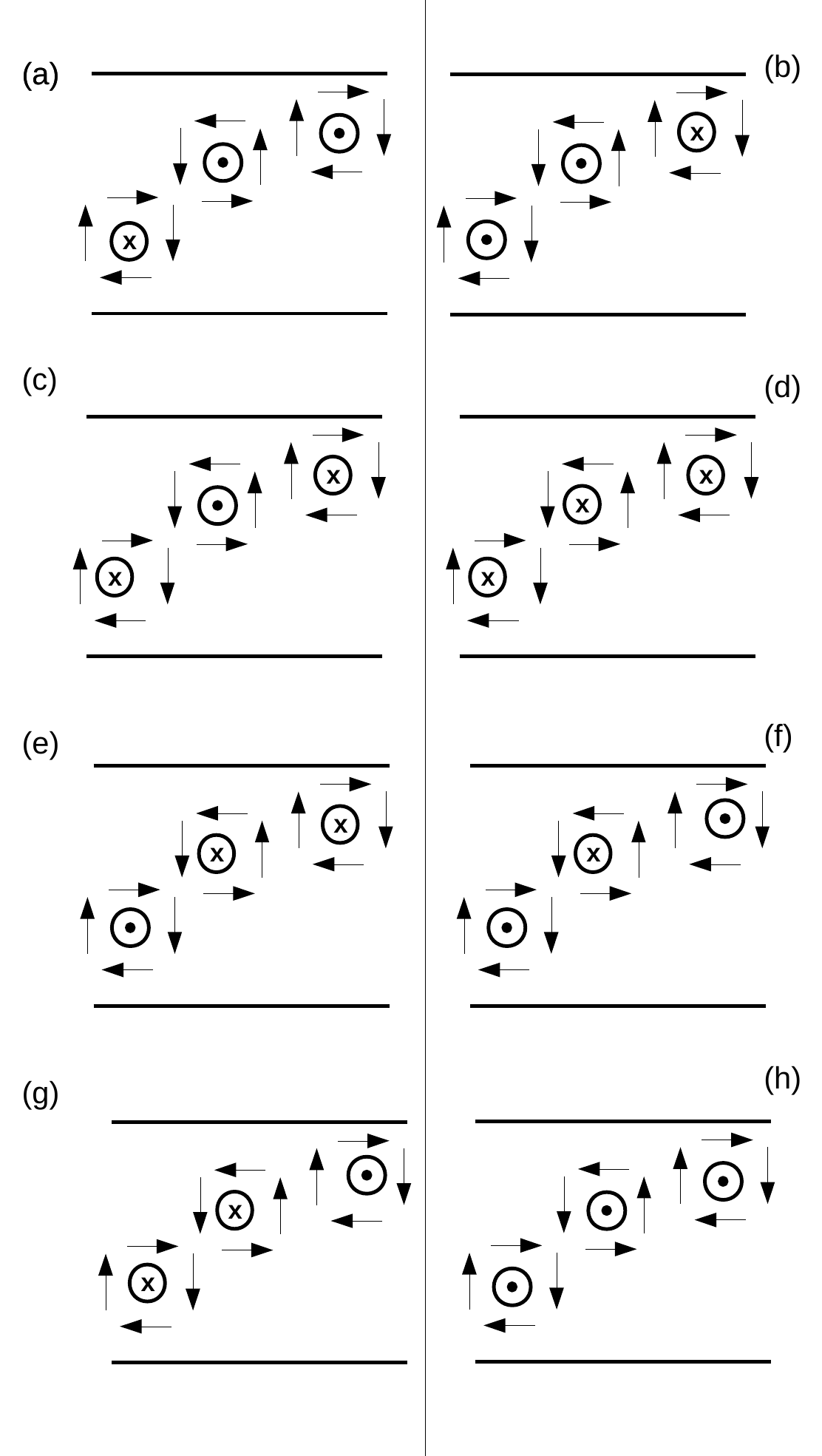}
\caption{(color online) Top view sketch of the different polarity configurations for a TVW structure.}
\label{fig:polarity3V}
\end{figure}

\begin{figure}[t!]
\leavevmode
\includegraphics[clip,width=0.5\textwidth]{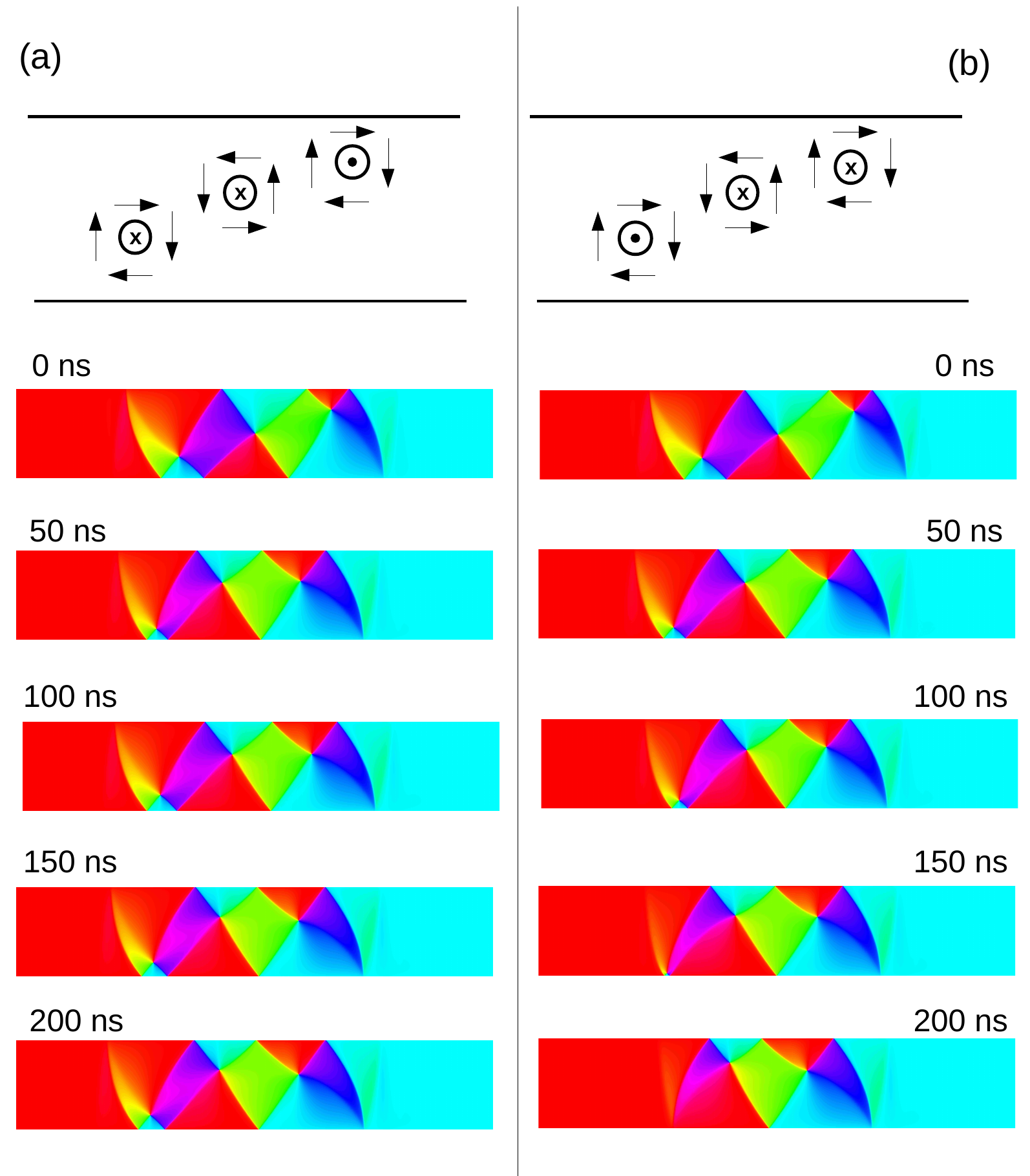}
\caption{(color online) TVW dynamics in a strip with $w=5120$ nm and 
$\Delta_z=25$ nm for $B_\text{ext}=0.05$ mT and two different configurations of the polarity. (a) TVW structure is stable against this field. (b) The initial TVW structure transforms into a DVW, which then exhibits steady dynamics within the simulation time scale.}
\label{fig:TVWtoDVW}
\end{figure}

\begin{figure}[t!]
\leavevmode
\includegraphics[clip,width=0.5\textwidth]{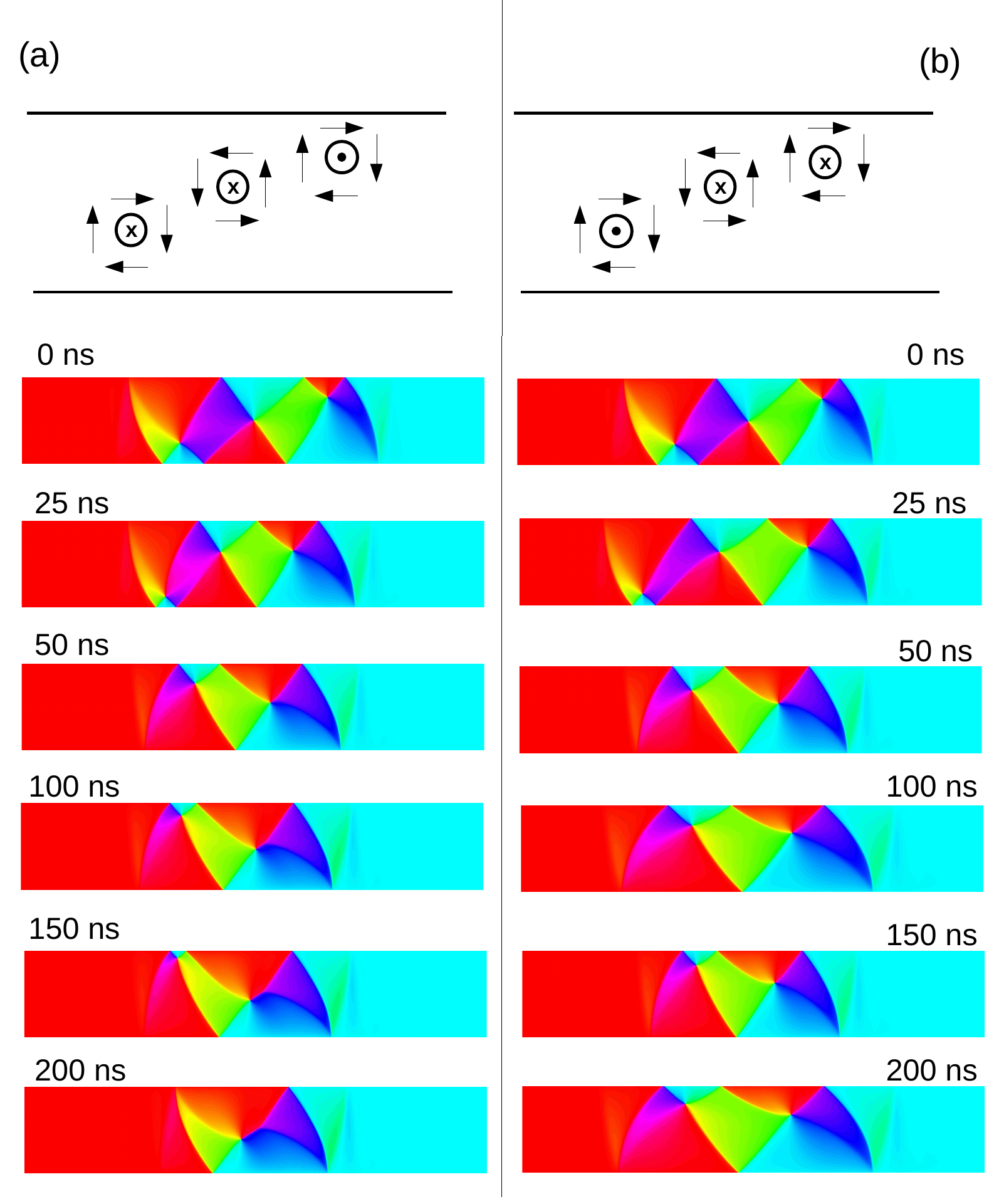}
\caption{(color online) TVW dynamics in a strip with $w=5120$ nm and 
$\Delta_z=25$ nm for  $B_\text{ext}=0.1$ mT and two different configurations of the polarity. (a) The initial TVW structure 
transforms first into a DVW, and then into a structure exhibiting the 
topology of a VW. (b) The initial TVW structure transforms into a DVW, which then exhibits steady dynamics within the simulation time scale.}
\label{fig:TVWtoDVWtoVW}
\end{figure}

\begin{figure*}[t!]
\leavevmode
\centering
\includegraphics[clip,width=1\textwidth]{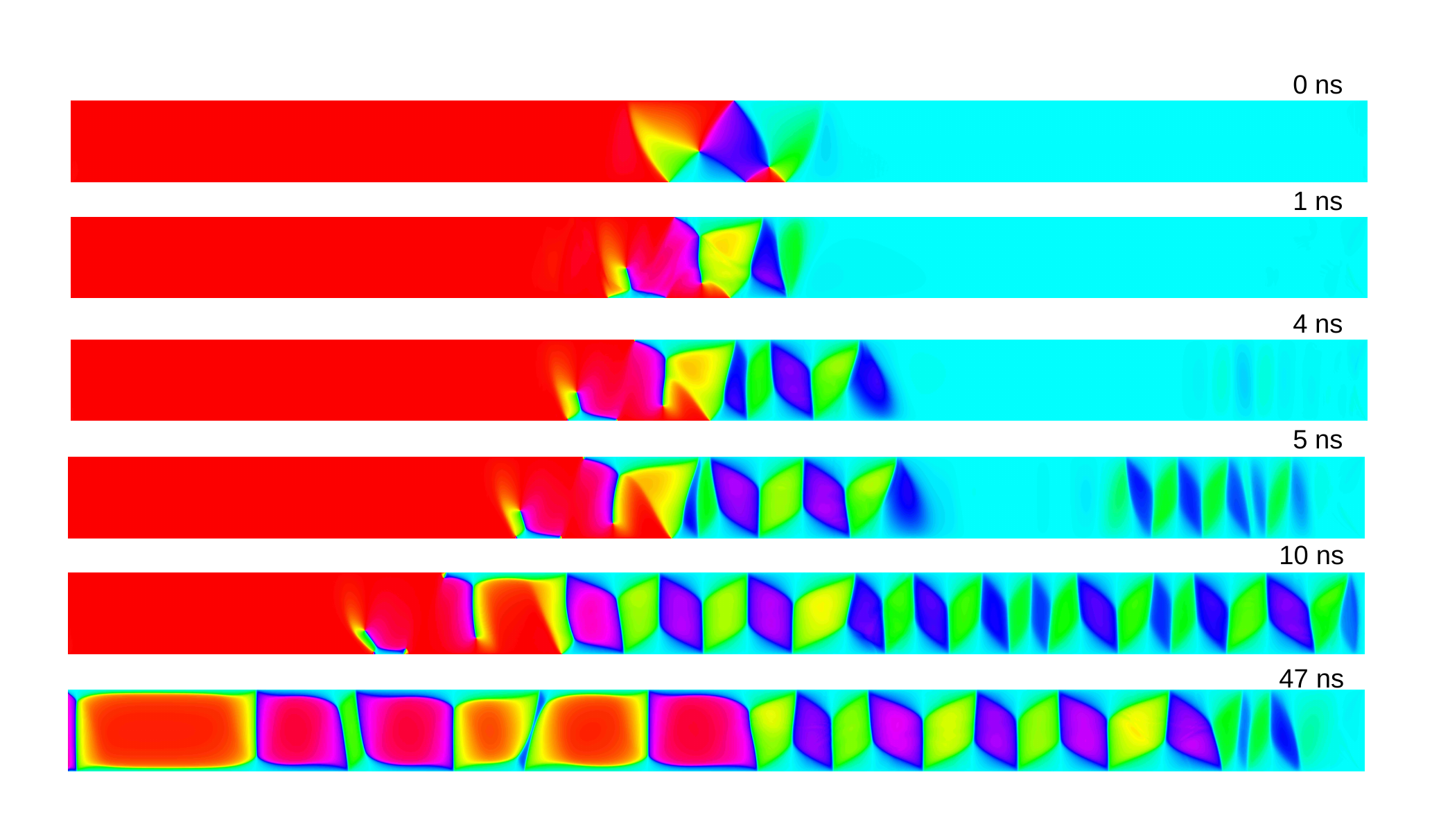}
\caption{(color online) Snapshots from a strip with $w=5120$ nm and
$\Delta_z=5$ nm with $B_\text{ext}=0.9$ mT. For these conditions the system
cannot support a compact DW, resulting in an instability where a complex
magnetization pattern fills the strip.}
\label{fig:fig11_dynamics}
\end{figure*}

For strips with even larger lateral dimensions, TVW is the equilibrium DW
structure\cite{nosotrosequilibrium}, see Fig.
\ref{fig:fig1_dynamics} (b). As in the case of DVW, also TVW tends
to transform into simpler DW structures under applied fields for topological 
reasons. TVWs are composed of three vortices and 6 HAV edge defects, see Fig.
\ref{fig:topological} (c). Thus, one or two vortices can be expelled from the strip with
relative ease, given that they can annihilate with two HAVs at the strip edge in
a process where the total winding number is conserved. 
Only in the case of very small fields the TVWs maintain their structure during motion depending on the polarities of the vortices. As in the DVW case, the polarity is fundamental for the stability of the TVW structure. As for TVWs there are three different vortices, the situation is more complex than in the case of DVWs. Fig. \ref{fig:polarity3V} shows the different combinations of the polarities of the TVW structure. Contrary to the DVWs, where for parallel polarity of the vortices the DW structure is stable at low fields, for TVW the sign of the polarity is also important. For a strip with $w=5120$ nm and $\Delta_z= 25$ nm under $B_\text{ext}=$ 0.05 mT, we have observed that when the three vortices are parallel with polarity $p=$ -1 [Fig. \ref{fig:polarity3V} (d)], the TVW is stable, whereas for polarity $p=$ +1 [Fig. \ref{fig:polarity3V} (h)] TVW transforms into a DVW (not shown). This is related to the gyrotropic force exerted on the vortices and the interactions between them. Depending on the polarity the gyrotropic force acts in one direction or the other, facilitating or preventing the expulsion of the vortex  out of the strip.

On the other hand, for antiparallel polarity also the position of the vortices with a given polarity is essential in the TVW dynamics. In Figs. \ref{fig:polarity3V} (e) and (g) two different polarity configurations are shown, where two vortices have polarity $p=$ -1 and the third one has $p=$ +1. These two configurations show different DW dynamics. To illustrate this we consider a strip  with $w=5120$ nm and $\Delta_z=25$ nm,
where TVW is one of the equilibrium DW structures\cite{nosotrosequilibrium}, considering both polarity configurations. For $B_\text{ext}$ = 0.05 mT and one of the above mentioned configuration, TVW is stable, keeping its structure during the simulation time, see Fig. \ref{fig:TVWtoDVW} (a). For the second one, 
the initial TVW transforms into a DVW as one of the three vortices is expelled from the strip. The resulting DVW exhibits steady dynamics within the simulation time scale, see Fig. \ref{fig:TVWtoDVW} (b). This is because the vortex expelled from the strip is the one with opposite polarity, and thus the two remaining vortices in the strip are parallel. 

To understand why one polarity configuration is stable and the other not, we have to consider the directions of movement of the vortices when a external magnetic field is applied. For the first configuration [Fig. \ref{fig:TVWtoDVW} (a)], due to the gyrotropic force, the vortices situated at the left and in the center move to the top strip edge, while the third one moves towards the bottom edge. The vortices of the ends of the DW structure are very far from the edges towards which they are moving, and the vortex at the middle cannot be expelled due to the chirality of the vortices. If the vortex at the center was expelled, the two remaining vortices had equal sense of rotation, and  the resulting DW structure would not respect the basic principles of topological defects in DWs. To obtain a total winding number equal to zero,  new topological defects should be created with the corresponding cost in energy\cite{nosotrosequilibrium}. As a result, the TVW is stable at this magnetic field. For the other case [Fig. \ref{fig:TVWtoDVW} (b)], the vortex on the left moves to the bottom edge, whereas the other two move towards the top edge. Here, the vortices at the ends are very close to the edges, and in principle we could expect that any of them could be expelled. However, as the vortex at the center also tries to go out, a competition between this vortex and the one situated at the right appears, keeping both inside the strip. Thus, the vortex on the left is expelled more easily.

As the applied field increases, the TVW transforms into simpler DW structures for all combinations of the vortex polarities. This can be clearly seen in Fig. \ref{fig:TVWtoDVWtoVW}, where the DW dynamics for the same strip with both polarity configurations are represented for $B_\text{ext}$ = 0.1 mT. For the configuration where the TVW is stable at a smaller field, now two of the three vortices are expelled, leading to a VW, see Fig. \ref{fig:TVWtoDVWtoVW} (a). Here the applied field is strong enough to push one of the vortices (the left one) out of the strip. The two remaining vortices have opposite polarities. As it was already seen in the previous section, the DVW with antiparallel polarity is unstable, and  the DW structure transforms into a VW. Fig. \ref{fig:TVWtoDVWtoVW} (b) shows that for the other configuration, the DW dynamics is similar to the one observed for $B_\text{ext}=$ 0.05 mT. TVW transforms in a DVW, which is stable during the simulation time because the polarities of the two vortices are parallel.

Similar results have been found for a larger strip with $w=6144$ nm and $\Delta_{z}=25$ nm (not shown). For this case, once two of the three vortices have been expelled, the remaining DW has the topology of a VW. However, this structure is more complex than typical VWs observed in more narrow strips. 
\\\\
Finally, we mention the fact that also for these very wide strips the system is,
for some combinations of the strip geometry and the applied field, unable
to support a compact DW. The related instabilities can be quite complex,
as illustrated in Fig. \ref{fig:fig11_dynamics} for a strip with 
$w=5120$ nm, $\Delta_z=5$ (corresponding to the phase boundary between
VW and DVW equilibrium structures\cite{nosotrosequilibrium}) and $B_\text{ext}=$ 
0.9 mT. In this case, a complex domain/magnetization pattern starts to
emanate from the initial DW, eventually filling the entire strip.
Thus, for such wide permalloy strips one starts to reach a situation resembling
increasingly that in thin films rather than nanostrips, where system-spanning
closure domain/Landau patterns are typically observed \cite{SHI-00,WAC-02,FIS-11}.

\section{SUMMARY AND CONCLUSIONS}
\label{summary}

To summarize, we have analyzed in detail DW dynamics driven by an external 
magnetic field in wide permalloy strips, by considering a set of example
geometries spanning a wide range of strip widths. Our main message is that
the system geometry crucially affects the DW dynamics also for lateral
strip dimensions beyond the typical nanostrip ones which have been extensively
studied in the literature. Already for relatively confined systems with
VW as the stable DW structure, several different dynamical behaviors are
obtained, including processes such as various periodic transitions between
different DW structures, the attraction-repulsion effect, etc. These behaviors are also 
visible in the profiles of the resulting $v(B_\text{ext})$ curves.
For wider strips  where the DVWs and TVWs
are the equilibrium DWs,  the polarities of the vortices play an important role in the dynamics. For small fields, DVW and TVW structures are stable, i.e. they maintain their structure depending on the polarity and the interactions between the vortices. As the size of the strip increases, for relatively
small fields in the range of a few mT, these structures tend to transform towards DW structures with less vortices. This effect is intimately linked to the topology of such
more complex equilibrium DW structures. The latter can be represented as composite
objects of topological defects, consisting of vortices and half-antivortex edge 
defects\cite{TCH-05}. The transitions towards simpler DW structures proceed via 
annihilation processes conserving the winding number, and thus the energy barrier 
for such transitions tends to be low. 

Another key feature of wide permalloy strips is that they are not always able
to support compact DWs. Indeed, as the strip width increases, shape
anisotropy becomes weaker, eventually leading to a situation where increasing
the DW width without limit becomes energetically favorable. In 
dynamical situations addressed here this tends to happen via different DW 
components acquiring different field-driven propagation velocities\cite{ZIN-11},
leading to gaps in the $v(B_\text{ext})$ curves as the DW and therefore
its velocity become ill-defined concepts. More generally, as the strip
width increases, one would expect a crossover to a thin film-like behavior
with system-spanning closure domain/Landau patterns\cite{SHI-00,WAC-02,FIS-11}
instead of localized DWs.

From an experimental perspective, our study highlights the importance of
performing simultaneous analysis e.g. by micromagnetic simulations, to verify
which of the large variety of complex dynamical behaviors corresponds to
the experimental conditions at hand. While our preliminary studies suggest that the attraction-repulsion effect is stable against adding random structural disorder\cite{leliaert2014} of moderate strength, future studies should address the effects
of e.g. thermal fluctuations and quenched disorder on DW dynamics in wide 
permalloy strips. For instance, the effect of e.g. edge roughness on the periodic
vortex core dynamics near the strip edge
would be interesting to address in detail.

\begin{acknowledgments}
We thank Ilari Rissanen for sharing computational resources. This work has been supported by the Academy of Finland through its Centres
of Excellence Programme (2012-2017) under project no. 251748, and via an Academy
Research Fellowship (LL, project no. 268302). We acknowledge the computational 
resources provided by the Aalto University School of Science ``Science-IT'' 
project, as well as those provided by CSC (Finland).
\end{acknowledgments}

\end{document}